\documentclass[acmtog]{acmart}
\acmSubmissionID{354}

\setcopyright{acmlicensed}
\acmJournal{TOG}
\acmYear{2022}
\acmVolume{41}
\acmNumber{6}
\acmArticle{261}
\acmMonth{12} 
\acmDOI{10.1145/3550454.3555484}

\setcitestyle{nosort,square}

\usepackage{microtype}      
\usepackage[capitalize]{cleveref}
\usepackage{xspace}         
\usepackage{mathtools}      
\usepackage{nicefrac,xfrac} 
\usepackage{overpic}        
\usepackage{pict2e}         
\usepackage{mathtools}

\usepackage{stmaryrd}  
\usepackage{newfloat}  
\usepackage{savesym}
\usepackage{algorithmicx}
\usepackage[noend]{algpseudocode} 
\usepackage{graphicx}
\usepackage{tikz}
\usepackage{grffile}  
\usepackage{soul}
\usetikzlibrary{arrows.meta}
\usepackage{wrapfig}
\usepackage{bbm}
\usepackage{flushend}
\usepackage[capitalize]{cleveref}
\usepackage{tabularx}
\usepackage{ctable}            
\usepackage{mathrsfs}
\usepackage[most]{tcolorbox}   
\usepackage[a]{esvect}           
\usepackage[outline]{contour}  
\usepackage{multirow}
\usepackage{wrapfig}
\usepackage{textcomp}    
\usepackage{enumitem}

\definecolor{Red}{rgb}{1,0.45,0.45}
\definecolor{Green}{rgb}{0.44,0.76,0.33}
\definecolor{Blue}{rgb}{0.46,0.7,1}
\definecolor{DarkGreen}{rgb}{0.0,0.6,0.0}
\definecolor{DomainColor}{gray}{0}
\definecolor{ExtSpaceColor}{rgb}{0.86,0,0.86}
\definecolor{ExtSpaceColorTwo}{rgb}{0.86,0.5,0.86}
\definecolor{TargetSpaceColor}{hsb}{0.45,0.87,0.81}
\definecolor{TargetMeasureColor}{rgb}{0.3,0.3,0.3}
\definecolor{RedClassColor}{hsb}{0.99,0.62,0.9}
\definecolor{BlueClassColor}{hsb}{0.61,0.6,0.9}
\definecolor{PurpleClassColor}{rgb}{0.86,0.35,0.88}
\definecolor{ProjectionColor}{rgb}{0.95,0.57,0.00}

\newcommand{\eg}{e.g.,\xspace}

\newcommand{\ie}{i.e.,\xspace}
\newcommand{\wrt}{w.r.t.\xspace}

\newcommand{\Paragraph}[1]{\paragraph{#1}}

\newcommand{\ScaleMath}[2]{\scalebox{#1}{\mbox{\ensuremath{\displaystyle #2}}}}

\newcommand{\mymath}[2]{\newcommand{#1}{\TextOrMath{$#2$\xspace}{#2}}}

\mymath{\Dim}{d}
\mymath{\Sphere}{\mathbb{S}}
\mymath{\SphereD}{{\Sphere^{\Dim-1}}}
\mymath{\ClassDim}{\mathcal{C}}
\mymath{\ClassSpace}{\mathcal{T}}
\mymath{\ClassID}{t}
\mymath{\Domain}{\mathcal{X}}
\mymath{\DomainY}{\mathcal{Y}}
\mymath{\DomainExt}{\overline{\Domain}}
\mymath{\Hypercube}{\mathcal{H}}
\mymath{\Reals}{\mathbb{R}}
\mymath{\RealsPlus}{\Reals^+}

\mymath{\Indicator}{\mathbf{1}}
\mymath{\dif}{\mathrm{d}}

\DeclareMathOperator*{\argmin}{arg\,min}
\mymath{\LebesgueMeasure}{\rho}

\mymath{\Class}{C}
\mymath{\ClassCoord}{c}
\mymath{\ClassPriority}{\kappa}
\mymath{\ClassFunc}{w}
\mymath{\PixelCount}{M}
\mymath{\ReferencePixel}{I}
\mymath{\ReferenceImage}{\boldsymbol{\ReferencePixel}}
\mymath{\EstimatedPixel}{Q}
\mymath{\EstimatedImage}{\boldsymbol{\EstimatedPixel}}
\mymath{\ErrorPixel}{\epsilon}
\mymath{\Error}{E}
\mymath{\ErrorImage}{\boldsymbol{\ErrorPixel}}
\mymath{\Point}{x}
\mymath{\PointExt}{\overline{x}}
\mymath{\PointSet}{X}
\mymath{\PointCount}{n}
\mymath{\PointSetExt}{\overline{\PointSet}}
\mymath{\Frequency}{m}
\mymath{\CDFInv}{F^{-1}_{\ProjectedTargetMeasure}}

\mymath{\Slice}{z}
\mymath{\ReconstFilter}{r}
\mymath{\PerceptFilter}{p}
\mymath{\VariationF}{\textit{V}}
\mymath{\Discrepancy}{\textit{D}}

\mymath{\PointSetMeasure}{\chi}
\mymath{\PointSetExtMeasure}{\overline{\chi}}
\mymath{\OptMeasure}{\nu}
\mymath{\TargetMeasure}{\mu}
\mymath{\ProjectedTargetMeasure}{\TargetMeasure^\theta}
\mymath{\TargetBarycenter}{{\overline{\TargetMeasure}}}
\mymath{\TransportOp}{T}
\mymath{\ExpectationOp}{\mathbb{E}}

\mymath{\Var}{\text{Var}}

\mymath{\numClasses}{K}
\mymath{\classIndex}{i}
\mymath{\classCount}{t}

\usepackage[nolist]{acronym}
\begin{acronym}
\acro{HVS}{human visual system}
\acro{MC}{Monte Carlo}
\acro{QMC}{Quasi-Monte Carlo}
\end{acronym}

\tcbset{
    colback=white!98!black,
    colframe=white!75!black,
    boxrule=0.15mm,
    arc=0.3mm,
    boxsep=0pt,left=2pt,right=4pt,top=2pt,bottom=4pt
}

\savesymbol{Comment}
\algrenewcommand\algorithmicindent{3mm}
\DeclareFloatingEnvironment[
    name=Alg.,
    placement=tbhp,
    within=none,
]{pseudocode}
\crefname{pseudocode}{Alg.}{Algs.}
\Crefname{pseudocode}{Algorithm}{Algorithms}

\DeclareGraphicsExtensions{.ai,.pdf,.jpg,.png}

\newcommand{\WrapFigureTemplate}[5]{%
    \setlength{\columnsep}{#2}
    \begin{wrapfigure}{r}{#1\columnwidth}%
      \begin{center}%
        \vspace{#3}%
        #5
        \vspace{#4}%
      \end{center}%
    \end{wrapfigure}%
    \leavevmode%
}

\DeclareDocumentCommand{\Outlined}{ O{black} O{white} O{0.55pt} m }{%
    \contourlength{#3}
    \contour{#2}{\textcolor{#1}{#4}}%
}

\newcommand{\undefinecolor}[1]{\expandafter\let\csname\string\color@#1\endcsname\undefined}

\begin{document}


\title{Scalable multi-class sampling via filtered sliced optimal transport}

\author{Corentin Sala\"un}
\affiliation{%
  \institution{Max-Planck-Institut f{\"u}r Informatik}
  \city{Saarbr{\"u}cken}
  \country{Germany}
}
\email{csalaun@mpi-inf.mpg.de}

\author{Iliyan Georgiev}
\affiliation{%
  \institution{Autodesk}
  \country{United Kingdom}
}
\email{me@iliyan.com}

\author{Hans-Peter Seidel}
\affiliation{%
  \institution{Max-Planck-Institut f{\"u}r Informatik}
  \city{Saarbr{\"u}cken}
  \country{Germany}
}
\email{hpseidel@mpi-sb.mpg.de}

\author{Gurprit Singh}
\affiliation{%
  \institution{Max-Planck-Institut f{\"u}r Informatik}
  \city{Saarbr{\"u}cken}
  \country{Germany}
}  
\email{gsingh@mpi-inf.mpg.de}



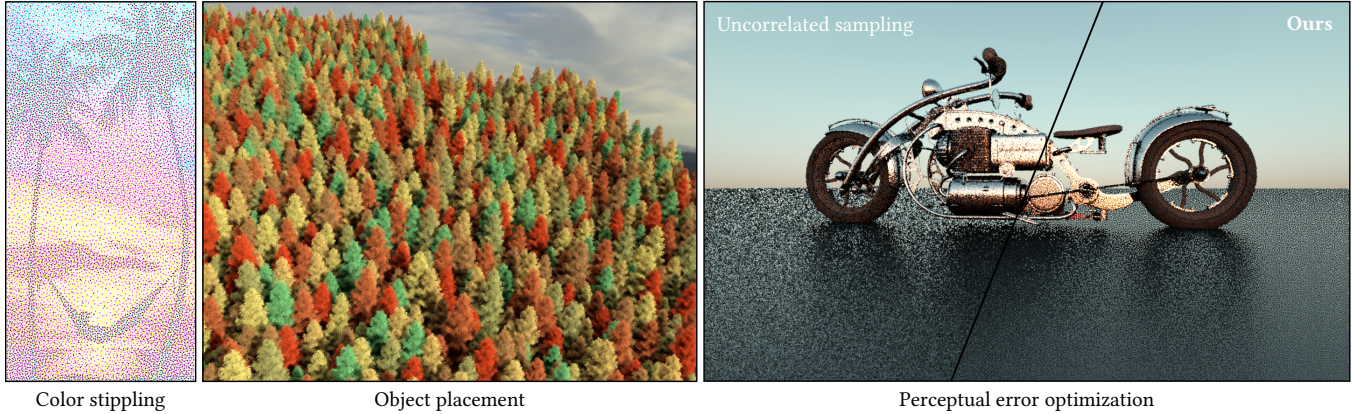
\begin{teaserfigure}

\newcommand{\OneImage}[1]{
    \begin{scope}
        \clip (0,0) -- (6.5,0) -- (6.5,5) -- (0,5) -- cycle;
        \path[fill overzoom image=figures/#1] (0,0) rectangle (6.5cm,5cm);
    \end{scope}
    \draw[line width=0.19mm] (0,0) -- (6.5,0) -- (6.5,5) -- (0,5) -- cycle;
}

\newcommand{\BeachImage}[1]{
    \begin{scope}
        \clip (0,0) -- (2.5,0) -- (2.5,5) -- (0,5) -- cycle;
        \path[fill overzoom image=figures/#1] (0,0) rectangle (2.5cm,5cm);
    \end{scope}
    \draw[line width=0.19mm] (0,0) -- (2.5,0) -- (2.5,5) -- (0,5) -- cycle;
}

\newcommand{\VerticalSplitTwoImages}[2]{
    \begin{scope}
        \clip (0,0)-- (0.0,4.5) -- (4.25,4.5) -- (4.25,0.0) -- cycle;
        \path[fill overzoom image=figures/#1] (0,0) rectangle (8.5,4.5);
    \end{scope}
    \begin{scope}
        \clip (4.25,0.0)-- (4.25,4.5) -- (8.5,4.5) -- (8.5,0.0) -- cycle;
        \path[fill overzoom image=figures/#2] (0,0) rectangle (8.5,s4.5);
    \end{scope}
    \draw[draw=white,thick] (3.25,4.5) -- (5.25,0.0);
}

\newcommand{\SlantedSplitTwoImages}[2]{
    \begin{scope}
        \clip (0,0)-- (0.0,5) -- (5.25,5) -- (3.25,0.0) -- cycle;
        \path[fill overzoom image=figures/#1] (0,0) rectangle (8.5,5);
    \end{scope}
    \begin{scope}
        \clip (3.25,0.0)-- (5.25,5) -- (8.5,5) -- (8.5,0.0) -- cycle;
        \path[fill overzoom image=figures/#2] (0,0) rectangle (8.5,5);
    \end{scope}
    \draw[line width=0.25mm,black] (5.25,5) -- (3.25,0.0);
    \draw[line width=0.19mm] (0,0) -- (8.5,0) -- (8.5,5) -- (0,5) -- cycle;
}

\small
\hspace*{-2.5mm}
\begin{tabular}{c@{\;}c@{\;}c@{}}
    \begin{tikzpicture}[scale=1.01]
        \BeachImage{teaser/beach_stippling_no_border.png}
    \end{tikzpicture}
        &
    \begin{tikzpicture}[scale=1.01]
        \OneImage{teaser/trees_teaser.jpg}
    \end{tikzpicture}
        &
    \begin{tikzpicture}[scale=1.01]
        \SlantedSplitTwoImages{teaser/chopper-titan_directlight_random_n1_gaussian_perceptual.jpg}{teaser/chopper-titan_directlight_Ours_n1_gaussian_perceptual.jpg}
        \begin{scope}
            \filldraw[white,ultra thick] (0.05, 4.93) circle (0pt) node[anchor=north west,rotate=0] 
            {Uncorrelated sampling \tiny};
            \filldraw[white,ultra thick] (8.4, 4.93) circle (0pt) node[anchor=north east,rotate=0] 
            {\textbf{Ours} \tiny};
        \end{scope}
    \end{tikzpicture}
    \\[-0.3mm]
    Color stippling & Object placement & Perceptual error optimization
\end{tabular}
  \vspace{-3mm}
    \caption{
        Demonstration of our multi-class sampling framework on three applications.
        Left: CMYK color stippling involves optimizing for 15 classes, each following a different, non-uniform density.
        Middle: 7 colors of trees and their union optimized jointly.
        Right: Distributing rendering error as blue noise, cast as a multi-class problem (4096 classes), showing improved visual fidelity over traditional uncorrelated-pixel sampling.
    }
    \label{fig:teaser}
    \vspace{2mm}
\end{teaserfigure}


\begin{abstract}

We propose a multi-class point optimization formulation based on continuous Wasserstein barycenters. Our formulation is designed to handle hundreds to thousands of optimization objectives and comes with a practical optimization scheme. We demonstrate the effectiveness of our framework on various sampling applications like stippling, object placement, and Monte-Carlo integration. We a derive multi-class error bound for perceptual rendering error which can be minimized using our optimization.
We provide source code at 
\url{https://github.com/iribis/filtered-sliced-optimal-transport}.

\end{abstract}



\maketitle

\section{Introduction}

Multi-class sampling finds numerous applications in computer graphics, such as object placement~\cite{wei2010multiclass}, visualization~\cite{ruizhen2020scatterplots,onzenoodt2021blue}, and multi-tone image stippling~\cite{secord2002weighted,schulz2021multiclass}. The goal of multi-class sampling is to produce a point set that satisfies multiple objectives simultaneously. An objective is to optimize a specific subset of points to follow a given target distribution. When the subsets are mutually disjoint, the task is relatively easy since each objective can be optimized separately. The difficulty arises in applications where the subsets overlap. Overlaps introduce conflicts between optimization objectives. A classical example is multi-tone image stippling, where individual color channels---each represented by a point subset, or \emph{class}---and their union(s) all have different target densities (\cref{fig:teaser}a). Such problems call for formulating a global optimization problem that can encode all objectives with a desired balance between them.

Existing multi-class solutions~\cite{wei2010multiclass,jiang2015bluenoise,qin2017wasserstein} do not scale to large numbers objectives, both in terms of means to specify many objectives and ability to optimize them in reasonable time and/or memory footprint. We propose a formulation based on continuous Wasserstein barycenters to achieve such scalability. Our formulation provides a simple way to specify multiple objectives at once and the desired balance between them. It is complemented by a gradient-descent optimization scheme that is only weakly sensitive to the number of objectives.

We demonstrate the utility of our framework on diverse applications that involve a large number of objectives, including color stippling, object placement, (progressive) Monte-Carlo integration, as well as blue-noise distribution of rendering error which we cast as a multi-class optimization problem. In summary:
\begin{itemize}[leftmargin=3.4mm]
    \setlength\itemsep{0.5mm}
\item 
    Our optimal-transport formulation allows specifying multiple optimization objectives at once via simple functions.
\item 
    Our stochastic optimization scheme scales to very large numbers of objectives.
\item 
    We derive an error bound for rendering error \wrt a given pixel-reconstruction kernel. When the kernel incorporates perceptual filtering, minimizing this bound yields point sets that distribute rendering error as blue noise over the image.
\end{itemize}

\section{Prior work}

In this section we review different sample correlations that are extensively studied in computer graphics. 

\subsection{Blue-noise sampling}

\citet{ulichney1987digital} pointed out that isotropic point distributions with predominantly high-frequency spectral content, namely \emph{blue noise}, cover the space evenly and look aesthetically pleasing. Since then blue-noise samples have been used for halftoning~\cite{ulichney1987digital}, object placement~\cite{kopf2006recursive,reinert2016projective}, stippling \cite{secord2002weighted,balzer2009capacity} and visualization~\cite{ruizhen2020scatterplots,onzenoodt2021blue}. In rendering, \citet{dippe85antialiasing} and \citet{cook86stochastic} also promoted samples with high-frequency content for anti-aliasing and image reconstruction. Various relaxation-based \cite{balzer2009capacity,deGoes2012blue,qin2017wasserstein}, tile-based~\cite{ostromoukhov2004penrose,ostromoukhov2007polyominoes,kopf2006recursive,wachtel2014fast} and number-theoretic-based methods~\cite{keller13quasi,abdalla2021optimizing} have since been proposed to generate high-quality blue-noise samples in multiple dimensions.

\Paragraph{Multi-class sampling}

Considering only the spatial locations of the samples could severely limits their applicability in real-world scenarios. Multi-class sampling allows incorporating non-spatial features while maintaining the blue-noise property of the spatial coordinates. \citet{wang1999properties} first showed the impact of multi-class sampling on colored halftoning of images. They developed a sampling algorithm that generates blue-noise quality in combinations of the R,G, and B channels of an image. \citet{wei2010multiclass} proposed multi-class sampling algorithm based on dart throwing. \citet{schmaltz2012multiclass} proposed electrostatic halftoning, whereas \citet{jiang2015bluenoise} used an SPH method to obtain multi-class samples. All these methods enforce multi-class blue noise through the use of an interaction matrix that encodes the spacing between class pairs. The matrix, however, can exhibit discontinuous changes in the off-diagonal entries, which represent the coupling between the different classes' distributions.\citet{chen2012variational} proposed a two-step algorithm based on capacity-constrained Voronoi tessellation to obtain a multi-class property. In the first step, each class is individually optimized, and in the next step their unions are optimized. \citet{chen2013bilateral} proposed a continuous multi-class sampling scheme limited to dart throwing and kernel-based optimization.

\citet{qin2017wasserstein} overcame these limitations via a multi-class framework based on optimal transport \cite{Rabin:Barycenter,Rachev:MassTransportation,Agueh:Barycenters}. Classes and their unions each have a target distribution. By optimizing for the Wasserstein barycenter of these measures they obtain a multi-class blue-noise point set. To handle conflicts between classes and to avoid regularity, they leveraged entropic regularization~\cite{cuturi2013sinkhorn}. That method works well but lacks the flexibility to specify many objectives. Their optimization also does not scale well with the number of objectives.
We propose a new formulation of multi-class sampling using sliced optimal transport which overcome these limitations and generalize multi-class sampling to different applications.

\subsection{Monte-Carlo integration}


In quasi-Monte Carlo literature, number-theoretic approaches are used to compute samples with good stratification, \ie low discrepancy~\cite{kuipers74uniform,niederreiter92random}. Discrepancy provides a measure of equidistribution and a bound for integration error via the Koksma-Hlawka inequality~\cite{ermakov2019koksmaHlawka}. Low-discrepancy point sets are widely used in image synthesis~\cite{keller13quasi,pharr2016physically}.


Following~\citet{durand2011frequency}, a theoretical connection has been established between the error in Monte-Carlo integration and the sampling power spectra~\cite{subr13fourier,pilleboue15variance,singh17convergence,singh2019analysis}. However, Fourier error remain insufficiently explored. Recently, \citet{paulin2020sliced} showen an error bound based on (sliced) optimal transport theory~\cite{pitie2005dimensional,villani2008optimal,bonneel2019spot,julien2011wasserstein}. That bound involves the Wasserstein distance~\cite{kantorovich1958space} which can be seen as the optimal-transport analog of the discrepancy metric. The samples obtained by minimizing this Wasserstein distance have blue-noise properties and compete with low-discrepancy distributions in terms of error.

\subsection{Perceptual error optimization}

Traditionally in Monte-Carlo rendering pixel values are estimated independently from one another, which yields white-noise distribution of error over the image. \citet{mitchell1991spectrally} noticed that error distributions with high-frequency power spectra give more pleasing appearance to noisy images. Following this observation, \citet{georgiev2016blue} proposed a dithering-inspired method to explicitly coordinate sampling across image pixels to achieve blue-noise error distribution. A number of follow-up works improved the quality and versatility of that basic approach \citet{heitz2019aLowDiscrepancy,abdalla2020screen,belcour2021rank1}, but all these methods are heuristic in nature and lack a formal treatment.

Recently, \citet{chizhov2022perceptual} adopted theory from halftoning to properly formalize perceptual error in rendering. We build on their formulation to derive a bound for this error; this bound can be minimized by our optimization scheme to produce sample sets that yield blue-noise error distribution.

\section{Preliminaries}

We begin our exposition by reviewing basic concepts in optimal transport theory upon which we will build our multi-class point-sampling formulation in \cref{sec:Theory}.

Optimal transport is concerned with moving the mass of one distribution to form another one~\citep{villani2008optimal,santambrogio2015optimal,peyre2018computationalOT}. Thinking of piles of sand, the question is what is the minimal cost (\ie total mass displaced per unit distance) required to transport the sand from an initial pile to a target pile. This cost gives a notion of \emph{distance} between two distributions.

\Paragraph{Wasserstein distance}

Formally, the optimal-transport distance between two measures (\ie distributions) \TargetMeasure and \OptMeasure is given by
\begin{equation}
    \label{eq:WassersteinDistance}
    W_p(\OptMeasure,\TargetMeasure) = \left(\, \inf_{\gamma \in \Gamma(\!\OptMeasure,\TargetMeasure)} \int_{\Domain^2} \|x - y\|^p \,\dif \gamma(x,y) \right)^{\nicefrac{1}{p}} \!\!\!\!\!,
\end{equation}
which is called the $p$-order Wasserstein metric~\cite{Book:OptimalTransport}. Here, $\|x - y\|$ denotes Euclidean distance on the domain $\Domain$. Intuitively, $\gamma$ is a transport plan (formally, a joint measure with marginals \TargetMeasure and \OptMeasure) such that $\dif \gamma(x,y)$ gives the (differential) amount of mass to be transported between any two points $x$ and $y$. The cost of doing so is thus $\|x - y\|^p \dif \gamma(x,y)$. In the space $\Gamma(\OptMeasure,\TargetMeasure)$ of all such plans, we are looking for one that minimizes the transportation cost over all pairs of points, \ie the integral in~\cref{eq:WassersteinDistance}.

Note that this formulation requires the two measures to have equal total mass, \ie $\OptMeasure(\Domain) = \TargetMeasure(\Domain)$. However, they do not necessarily have to be probability measures, \ie to have unit mass.

\Paragraph{Sliced Wasserstein distance}

Computing the optimal transport plan in the Wasserstein distance~\eqref{eq:WassersteinDistance} can be very costly. A variant that is generally easier to solve involves computing only one-dimensional distances over all possible line projections of the two measures~\citep{Pitie:SlicedOptimalTransport,Rabin:Barycenter,Bonneel:Sliced}:
\begin{align}
    \label{eq:SlicedWassersteinDistance}
    SW_p(\OptMeasure,\TargetMeasure)
        \, = \int_\SphereD \!\!\! W_p\big( \OptMeasure^\theta,\ProjectedTargetMeasure \big) \,\dif \theta
\end{align}
In this so-called sliced Wasserstein distance, $\theta \in \SphereD$ is a point on the $(\Dim\!-\!1)$-dimensional sphere, and $\OptMeasure^\theta$ and $\ProjectedTargetMeasure$ are the orthogonal projections onto the line through $\theta$ of the two measures.

\Paragraph{(Sliced) Wasserstein barycenter}

The Wasserstein distance provides an intuitive means to construct weighted averages of distributions, beyond simple mixtures (\ie density averages). Generalizing the notion of barycentric interpolation between points, the Wasserstein barycenter \OptMeasure interpolates between several measures $\TargetMeasure_i$~\citep{Agueh:Barycenters,Rabin:Barycenter}:
\begin{equation}
    \label{eq:WassersteinBarycenter}
    \OptMeasure = \argmin_{\OptMeasure} \sum_i \lambda_i W_p^p(\OptMeasure,\TargetMeasure_i),
\end{equation}
where the scalar weights $\lambda_i$ sum up to one. The Wasserstein barycenter can be seen as a means to compromise between the various objectives, here finding a distribution that minimizes the (weighted) distance to several targets. Replacing $W_p$ by its sliced variant $SW_p$ enables the practical computation of the barycenter through repeated 1D optimizations~\citep{Rabin:Barycenter,Bonneel:Sliced}.

\section{Multi-class optimal transport}
\label{sec:Theory}

\begin{figure}[t!]
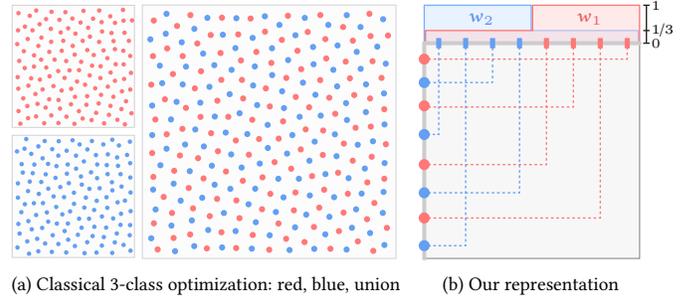

    \begin{overpic}[abs,unit=0.3mm,grid=false]{figures/formulation_example2}
        \footnotesize
        \put(250,106){\color{RedClassColor}$\ClassFunc_{\,1}$}
        \put(202,106){\color{BlueClassColor}$\ClassFunc_{\,2}$}
        \put(283.5,111){\tiny$1$}
        \put(283.5,100.5){\tiny$\nicefrac{1}{3}$}
        \put(283.5,94.4){\tiny$0$}
        \put(0,-13){\fontsize{7.3}{7}\selectfont (a)~Classical 3-class optimization: red, blue, union}
        \put(191,-13){\fontsize{7.3}{7}\selectfont (b)~Our representation}
    \end{overpic}
    \vspace{1mm}
    \caption{
        (a)~A classical 3-class example, where the red and blue point subsets represent one uniform-density objective (class) each and their union is another objective, all three with equal optimization priority.
        (b)~We represent this three-objective problem using two staircase functions, one per color, defined on an extra dimension (\eg the point indices). The overlap between the functions implicitly specifies the third (union) objective.
    }
    \label{fig:FormulationExample}
\end{figure}

Multi-class sampling involves producing a point set \mbox{$\PointSet = \{\Point_i\}_{i=1}^\PointCount \!\subset \Domain$} with a unique optimization objective for each of multiple subsets. For example, in~\cref{fig:FormulationExample}a the goal is to achieve high-quality isotropic uniform distribution for each point color and their union. We thus have three optimization objectives.

Typically, objectives are specified individually, which is reasonable when their number is small as in the applications considered by prior works~\citep{wei2010multiclass,jiang2015bluenoise,qin2017wasserstein}. However, this approach does not scale to our goal of handling large numbers of objectives. We want a principled and convenient means to specify objectives in bulk and to manage the conflicts between them. We also seek a more abstract way to specify point subsets beyond simple indices, to enable applications where points are more naturally grouped by other attributes.

In this section, we propose a novel formulation of the multi-class sampling problem based on optimal transport, to achieve the aforementioned goals. We operate on an extended space where the extra dimension is used for point classification and the remaining dimensions are optimized. \Cref{fig:FormulationExample}b shows a simple example where we represent the three optimization objectives in \cref{fig:FormulationExample}a using \emph{two} functions on that extra dimension. In the remainder of this section we introduce each component of this figure. \Cref{tab:Notation} summarizes the most important notations we use. 

\subsection{Classes and subclasses}
\label{subsec:ClassesAndSubclasses}

A class is an optimization objective specified by subset of points and a target distribution. The subset is typically given as a range of indices. We begin by generalizing the index space. Specifically, we extend the optimization space \Domain by a \emph{classification dimension} \ClassDim. Given a point set \PointSet, a corresponding extended point set $\PointSetExt = \{\PointExt_i\}_{i=1}^\PointCount$ is created by distributing \PointCount class coordinates $\ClassCoord_i$ uniformly in \ClassDim and assigning them arbitrarily to the optimization points: $\PointExt_i = (\ClassCoord_i, \Point_i)$.

\Cref{fig:Classes}a illustrates this setup. The best choice for \ClassDim depends on the application. \ClassDim can be multi-dimensional but most often it will simply be the unit line, \ie $\ClassDim = [0,1]$, and the class coordinates will be the normalized indices in the base point set: $\PointExt_i = (i/\PointCount, \Point_i)$. This normalization will allow us to define classes directly on the (fixed) extended space $\ClassDim \times \Domain$, independently of (the size of) any particular point set. Sometimes a different class dimension is more natural, \eg in rendering-error minimization \ClassDim will be the 2D image plane~(see \cref{sec:ErrorOptimization}).

\Paragraph{Classes}

The classification dimension \ClassDim is not part of the optimization and is invariant to the number of points to be optimized. This allows to isolate point subsets by taking subregions of \ClassDim, as illustrated in \cref{fig:Classes}b. One way to specify a subregion is through the support of some function \ClassFunc on \ClassDim; the figure shows the simplest case of a box function. 
The isolated points can then be optimized toward a target distribution~\TargetMeasure.

We can now give a concrete definition of a \emph{class} as a pair~$(\ClassFunc,\TargetMeasure)$. For a specific point set \PointSetExt, the optimization objective is to have all points within the support of \ClassFunc follow the distribution \TargetMeasure.

\begin{figure}[t!]
    \begin{overpic}[abs,unit=0.25mm,grid=false]{figures/classes}
        \footnotesize
        \put(-4,-9.3){\color{DomainColor}$\Domain$}
        \put(103,97){\color{ExtSpaceColor}$\ClassDim$}
        \put(64.5,5){\fontsize{6.7}{7}\selectfont${\color{black!60!white}\PointExt \! = \! ({\color{ExtSpaceColor}\ClassCoord}, {\color{black}\Point})}$}
        \put(178,108){\color{ExtSpaceColor}$\ClassFunc$}
        \put(109.5,29){\color{TargetMeasureColor}$\TargetMeasure$}
        \put(233.5,112.5){$\Slice$}
        \put(343,117.5){\tiny\selectfont $1$}
        \put(343,108.5){\tiny$0.5$}
        \put(343,98){\tiny\selectfont $0$}
        \put(301.5,105){\color{ExtSpaceColor}\scalebox{0.9}{$\ClassFunc \! > \! \Slice$}}
        \put(241,66){\color{ExtSpaceColor}\scalebox{0.75}{$\PointSetExt_{\!\ClassFunc > \Slice} \! \subseteq \PointSet$}}
        \put(227.5,29){\color{TargetMeasureColor}$\TargetMeasure$}
        \put(8,-14){\fontsize{7}{7}\selectfont (a)~Point-set extension}
        \put(133,-14){\fontsize{7}{7}\selectfont (b)~A simple class}
        \put(233,-14){\fontsize{7}{7}\selectfont (c)~A class with subclasses}
    \end{overpic}
    \vspace{-1mm}
    \caption{
        (a)~We extend the dimension of a point set by assigning to each point $\Point$ a unique class coordinate $\ClassCoord \in \ClassDim$ that remains fixed during optimization. This coordinate will often be the (normalized) index of the point.
        (b)~A subregion in dimension $\ClassDim$ isolates a subset of points that can be optimized to follow a target distribution $\TargetMeasure$. The subregion can be given by the support of a function $\ClassFunc$ on $\ClassDim$, here a box function. We define a class as a pair $(\!\ClassFunc,\TargetMeasure)$.
        (c)~A non-trivial class function yields multiple subclasses (sharing a target distribution) enumerated by slicing $\ClassFunc$.
        The staircase function here yields two subclasses, selecting all ({\color{ExtSpaceColorTwo}$\PointSetExt_{\!\ClassFunc > 0}$}) and half ({\color{ExtSpaceColor}$\PointSetExt_{\!\ClassFunc > 0.5}$}) the class' points.
    }
    \label{fig:Classes}
\end{figure}

\Paragraph{Subclasses}
\label{par:Subclasses}

The expressiveness of our formulation comes from the use of non-trivial (\ie non-box) \emph{class functions} \ClassFunc. Such a function can specify multiple optimization objectives. \Cref{fig:Classes}c shows a simple staircase function. Two unique intervals on \ClassDim can be extracted by thresholding that function, selecting all ($\ClassFunc > 0$) and half ($\ClassFunc > 0.5$) of the class' points, respectively. Each such interval represents a distinct optimization objective.

Extending the example to a more complex staircase---or even smooth---function \ClassFunc, allows us to specify an arbitrary number of
\WrapFigureTemplate{0.26}{5.5mm}{-2.3mm}{-5.7mm}{%
    \begin{overpic}[abs,unit=0.25mm,scale=1,grid=false]{figures/smooth-function}
        \put(-7,-1){\small $0$}
        \put(-7,26.2){\small $\Slice$}
        \put(-7,47){\small $1$}
        \put(32.5,31.5){\color{ExtSpaceColor}\scalebox{0.77}{$\ClassFunc \! > \! \Slice$}}
    \end{overpic}%
}%
sub-objectives, or \emph{subclasses}, all sharing the target distribution \TargetMeasure. Each subclass selects a point subset within the support of the thresholded, or \emph{filtered}, class function \ClassFunc at a value~$\Slice \geq 0$ (see the inline figure). Formally, the support of the filtered function is the set \mbox{$\{\ClassCoord \colon \ClassFunc(\ClassCoord) \!>\! \Slice\} \subseteq \ClassDim$}. A~subclass is then defined by a tuple~$(\ClassFunc, \TargetMeasure, \Slice)$.

\ctable[
    caption = {
        List of notations used throughout the document, which are also illustrated graphically in \cref{fig:Classes,fig:Formulation,fig:StochasticMinimization}. For simplicity, we denote point sets and their corresponding Dirac point-mass measures using the same symbol~$\PointSet$.
    },
    captionskip = -2mm,
    label = tab:Notation,
    width = \columnwidth,
    pos = t
]{ll}{}{
    \FL
    \!\!\textbf{Symbol}\!\! & \textbf{Description}
    \ML
    \!\!$\Domain, \ClassDim, \DomainExt$\!\! & Sampling domain, class domain, $\DomainExt = \ClassDim \times \Domain$
    \NN
    \!\!$\Point, \ClassCoord, \PointExt$ & Sample point, class coordinate, extended point $\PointExt \! = \! (\ClassCoord, \Point)$
    \NN
    \!\!$\PointSet, \PointSetExt$ & Point set $\{ \Point_i \}_{i=1}^\PointCount \! \subset \Domain$, extended point set $\{ \PointExt_i \}_{i=1}^\PointCount \! \subset \DomainExt$
    \NN
    \!\!$\ClassFunc, \TargetMeasure$ & Class function on $\ClassDim$, target distribution on $\Domain$
    \NN
    \!\!$\ClassSpace, \ClassFunc_\ClassID, \TargetMeasure_\ClassID$\!\!\! & Space $\ClassSpace \! = [0,1] \! \ni \! \ClassID$ of classes $(\ClassFunc_\ClassID, \TargetMeasure_\ClassID)$
    \NN
    \!\!$\PointSetExt_{\ClassFunc > z}$ & Subset extraction via filteringc: \scalebox{0.92}{$\{ \Point \colon \! (\ClassCoord,\Point) \! \in \! \PointSetExt, \ClassFunc(\ClassCoord) \! > \! z \} \! \subseteq \! \PointSet$}
    \NN
    \!\!$\TargetMeasure_{\ClassFunc > z}$ & Scaling $\TargetMeasure$ to match its mass to that of subset $\PointSetExt_{\ClassFunc > z}$
    \NN
    \!\!$\PointSet^\theta, \ProjectedTargetMeasure$ & Projections of distributions onto axis $\theta$
    \LL
}

\Paragraph{Point-set filtering}

A class is defined on a continuous extended space \DomainExt. 
A smooth class function defined in \DomainExt thus specifies an entire continuum of subclasses. For a specific point set \PointSetExt, their effective count is capped by the number of points. The subset of points selected by a subclass $(\ClassFunc, \TargetMeasure, \Slice)$ is obtained via a filtering operation: $\PointSetExt_{\ClassFunc > \Slice} \subseteq \PointSet$ contains all points $\Point_i \in \PointSet$ for which $w(c_i) > \Slice$ (see \cref{fig:Classes}c). Note that this also strips the class coordinate, producing a subset of the original point set $\PointSet$. A threshold value $\Slice=0$ selects all points in a class, and larger $\Slice$ values yield smaller subsets. The example in \cref{fig:FormulationExample}b demonstrates one such example with two staircase functions. For each staircase function, when \ClassFunc > 1/3 it selects half of the points, otherwise for \ClassFunc > 0 all points are selected.

\subsection{Subclass barycenter}
\label{subsec:SubclassBarycenter}

The objective specified by one subclass can be satisfied by minimizing the Wasserstein distance between the corresponding point subset and the target. However, subclasses overlap, meaning that a point can be subject to multiple objectives. A compromise between all subclass objectives can be achieved via a barycenter that minimizes all involved Wasserstein distances simultaneously. For a continuous class function \ClassFunc, the barycenter takes an integral form:
\begin{align}
    \label{eq:SubclassBarycenter}
    \PointSetExt
        \,=\, \argmin_{\PointSetExt}
        \begingroup
            \color{gray}
            \underbracket[0.5pt][1.7pt]{\color{black}\int_\Reals W_p^p\big(\,\PointSetExt_{\ClassFunc > \Slice},\TargetMeasure_{\ClassFunc > \Slice} \big) \,\dif \Slice}_{B_p(\PointSetExt, \ClassFunc, \TargetMeasure)}
        \endgroup
        \,,
\end{align}
which is a continuous variant of \cref{eq:WassersteinBarycenter}, with the difference that we have one target distribution $\TargetMeasure$ and multiple optimization (point) distributions. We scale the target distribution, $\TargetMeasure_{\ClassFunc > \Slice}$, to match the total mass of the subclass $\PointSetExt_{\ClassFunc > \Slice}$. The scaling factor is the relative
\WrapFigureTemplate{0.24}{6.5mm}{-2.2mm}{-8.4mm}{
    \begin{overpic}[abs ,unit=0.25mm,scale=1]{figures/step-function}
        \put(-9.4, -1){\small $0$}
        \put(-10.4,13){\small $\Slice_1$}
        \put(-10.4,34){\small $\Slice_2$}
        \put(-10.4,53){\small $\Slice_3$}
        \put(-9.6, 62){\small $1$}
    \end{overpic}
}%
number of points in the subclass compared to the size of the entire point set. When the class function \ClassFunc is piecewise constant, with levels $0 = \Slice_0, \Slice_1, \ldots, \Slice_s = 1$ (see inline figure), the integral becomes a sum, turning the problem into a discrete barycenter:
\begin{align}
    \label{eq:SubclassBarycenterDiscrete}
    \PointSetExt
        &= \argmin_{\PointSetExt} \sum_{j=1}^s \lambda_j W_p^p\big(\,\PointSetExt_{\ClassFunc > \Slice_j},\TargetMeasure_{\ClassFunc > \Slice} \big),
    \;\;\text{with}\;\;
    \lambda_j = z_j \! - \! z_{j-1}.\!
\end{align}
We enforce $\ClassFunc$ to have a maximum value of one to ensure that the weights $\lambda_j$ sum up to unity. For a trivial (box-function) class, the barycenter simplifies to the single objective of minimizing the Wasserstein distance between all class points and the target $\TargetMeasure$.

Note that the class function \ClassFunc encodes both the shape and the relative importance of each subclass (\ie its weight $\lambda_i$ in the discrete case). A class $(\ClassFunc, \TargetMeasure)$ thus completely describes an entire optimization problem~\eqref{eq:SubclassBarycenter}, independently of the point-set size.

\subsection{Multi-class barycenter}
\label{subsec:MultiClassBarycenter}

While a single subclass barycenter can completely describe some optimization tasks, it is not sufficiently expressive to model many practical problems. For example, having overlapping point subsets follow different target distributions. Even with one target, multiple subsets can be assembled into a single class only if they are nested into one another. The example in  \cref{fig:FormulationExample} cannot be modelled with a single class as the red and blue subsets are disjoint. Such problems require specifying and optimizing across multiple classes.

\Paragraph{Continuous case}

\begin{figure}[t!]
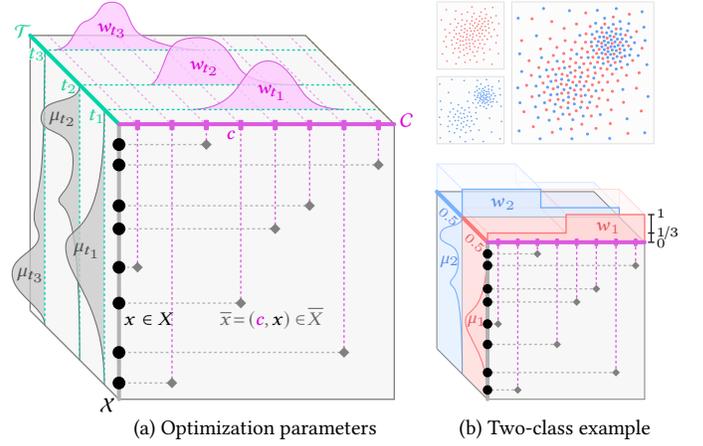

    \begin{overpic}[abs,unit=0.25mm,grid=false]{figures/formulation}
        \footnotesize
        \put(46.5,-2){\color{DomainColor}$\Domain$}
        \put(206,150){\color{ExtSpaceColor}$\ClassDim$}
        \put(1,197){\color{TargetSpaceColor}$\ClassSpace$}
        \put(41,153){\color{TargetSpaceColor}$\ClassID_1$}
        \put(26,169){\color{TargetSpaceColor}$\ClassID_2$}
        \put(9,186){\Outlined[TargetSpaceColor][white]{$\ClassID_3$}}
        \put(33,83){\color{TargetMeasureColor}$\TargetMeasure_{\ClassID_1}$}
        \put(20,153){\color{TargetMeasureColor}$\TargetMeasure_{\ClassID_2}$}
        \put(3,70){\color{TargetMeasureColor}$\TargetMeasure_{\ClassID_3}$}
        \put(130,167){\color{ExtSpaceColor}$\ClassFunc_{\ClassID_1}$}
        \put(94,180){\color{ExtSpaceColor}$\ClassFunc_{\ClassID_2}$}
        \put(45,200){\color{ExtSpaceColor}$\ClassFunc_{\ClassID_3}$}
        \put(60,44){$\Point \in \PointSet$}
        \put(111,44){${\color{black!60!white}\PointExt \! = \! ({\color{ExtSpaceColor}\ClassCoord}, {\color{black}\Point}) \!\in\! \PointSetExt}$}
        \put(114.5,143){\color{ExtSpaceColor}$\ClassCoord$}
        %
        \put(242.5,45){\scriptsize\color{RedClassColor}$\TargetMeasure_1$}
        \put(228,77){\scriptsize\color{BlueClassColor}$\TargetMeasure_2$}
        \put(310,94){\color{RedClassColor}$\ClassFunc_{\,1}$}
        \put(254,107){\color{BlueClassColor}$\ClassFunc_{\,2}$}
        \put(240.5,90){\tiny\color{RedClassColor}\rotatebox{317}{$0.5$}}
        \put(226.5,104){\tiny\color{BlueClassColor}\rotatebox{317}{$0.5$}}
        \put(343,101){\tiny$1$}
        \put(343,91.3){\tiny$\nicefrac{1}{3}$}
        \put(343,85){\tiny$0$}
        %
        \put(65,-14){\fontsize{8}{7}\selectfont (a)~Optimization parameters}
        \put(238,-14){\fontsize{8}{7}\selectfont (b)~Two-class example}
    \end{overpic}
    \vspace{0.1mm}
    \caption{
        (a)~Our continuous optimization formulation yields a barycenter between classes $(\!\ClassFunc_\ClassID,\TargetMeasure_\ClassID)$, each identified by point \ClassID on the unit line \ClassSpace; we show three classes here. Smooth class-function falloffs allow for accurate control over conflicts resulting from overlaps.
        (b)~A classical, fully discrete example with non-uniform target distributions. Each class function assigns equal optimization priority to each half of the points and their union. The top shows an optimized 256-point set.
    }
    \label{fig:Formulation}
\end{figure}

To specify multiple classes, we add one more dimension to our representation from \cref{fig:Classes}, illustrated in \cref{fig:Formulation}a. Each point \ClassID on the \ClassSpace axis identifies a class $(\ClassFunc_\ClassID,\TargetMeasure_\ClassID)$. \ClassSpace can be multi-dimensional but for simplicity we use the unit line: $\ClassSpace = [0,1]$. The different classes generally have conflicting objectives due to overlaps in their associated functions $\ClassFunc_\ClassID$. As discussed in~\cref{subsec:SubclassBarycenter}, the compromise between these objectives can be expressed as the Wasserstein barycenter
%
\begin{equation}
    \label{eq:MultiClassBarycenter}
    \PointSetExt \,=\, \argmin_{\PointSetExt} \int_0^1 \! \!\!
    \begingroup
        \color{gray}
        \underbracket[0.5pt][1.7pt]{\color{black}\int_0^1 \!\! W_p^p\big(\,\PointSetExt_{\ClassFunc_\ClassID>\Slice},\TargetMeasure_{\ClassID,\ClassFunc_\ClassID>\Slice} \big) \,\dif \Slice}_{\qquad B_p(\PointSetExt, \ClassFunc_\ClassID, \TargetMeasure_\ClassID) \;\;\; \text{\cref{eq:SubclassBarycenter}}}
    \endgroup
    \dif \ClassID
\end{equation}
%
across all classes (outer integral) and their subclasses (inner integral), recalling that our class functions have a maximum value of one.

\Paragraph{Discrete case}

Not every identifier $\ClassID$ has to map to a unique class. When the classes are a finite number $n$, the mapping is piecewise constant: $0 \! = \! \ClassID_0, \ClassID_1, \ldots, \ClassID_n \! = \! 1$, and every $\ClassID \! \in \! [t_{i-1}, t_i)$ maps to the class $(\ClassFunc_i, \TargetMeasure_i)$. In the fully discrete case, where each class has a staircase-like function, the optimization problem~\eqref{eq:MultiClassBarycenter} becomes a sum:
\begin{align}
    \label{eq:MultiClassBarycenterDiscrete}
    \PointSetExt \,=\, \argmin_{\PointSetExt} \sum_{i=1}^n \sum_{j=1}^{s_i} \ClassPriority_i \lambda_{i,j} W_p^p\big(\,\PointSetExt_{\ClassFunc_i > \Slice_{i,j}},\TargetMeasure_{i,\ClassFunc_i > \Slice_{i,j}} \big),
\end{align}
where $\ClassPriority_i = \ClassID_i  -  \ClassID_{i-1}$ are the class weights, and $\lambda_{i,j} = z_{i,j} - z_{i,j-1}$ are the subclass weights as in \cref{eq:SubclassBarycenterDiscrete}.

\Cref{fig:Formulation}b extends the example from \cref{fig:FormulationExample} to non-uniform target distributions. We formalize this optimization problem using only two classes: ($\ClassFunc_1,\mu_1$) and ($\ClassFunc_2,\mu_2$). Filtering the point set using these class functions give four subsets: $\{\PointSetExt_{\ClassFunc_1 > 0}, \PointSetExt_{\ClassFunc_1 > \nicefrac{1}{3}},\PointSetExt_{\ClassFunc_2 > 0}, \PointSetExt_{\ClassFunc_2 > \nicefrac{1}{3}}\}$. \Cref{eq:MultiClassBarycenterDiscrete} aims to find the barycenter defined by the Wasserstein distance wrt each subset. It is important to note that the target distribution is only defined for the red ($\mu_1$) and the blue points ($\mu_2$) and not their union. The union will be aiming towards a barycenter of $\mu_1$ and $\mu_2$ since each class function considers all the points (the union) when $\ClassFunc_i > 0$.

\Paragraph{Discussion}

Note that the class functions \ClassFunc (defined along the magenta axis in~\cref{fig:Formulation}) can overlap. Overlaps allow increasing the ``footprints'' of individual objectives to target more points than would be otherwise possible; however, they also introduce conflicts. In regions of overlap, the values of each function indicate its class' relative local optimization priority. Consequently, using functions with smooth falloffs allows us to precisely control the barycentric trade-off between class objectives in such regions. In the general case of diverse targets $\TargetMeasure_1$ and $\TargetMeasure_2$, the class overlaps can make it difficult to satisfy simultaneously the objectives. Generally, we want to avoid having two classes assign high priority to the same region. Class functions should ideally be arranged to overlap only in their tails; this helps better satisfy each class' objective by minimizing conflicts and reducing optimization pressure. Class functions can be designed depending on the application.

Our more traditional-looking discrete barycenter~\eqref{eq:MultiClassBarycenterDiscrete} makes it
%
\WrapFigureTemplate{0.34}{4.0mm}{-3.5mm}{-3.0mm}{%
    \begin{overpic}[abs,unit=0.25mm,scale=1,grid=false]{figures/3vs2class}
        \put(9,51){\small 2-class configuration}
        \put(111,79){\tiny$1$}
        \put(111,67){\tiny$\nicefrac{1}{3}$}
        \put(111,60.5){\tiny$0$}
        \put(111,28){\tiny$1$}
        \put(111,9){\tiny$0$}
        \put(9,0){\small 3-class configuration}
    \end{overpic}%
 }%
clear that the atomic optimization objective in our framework is the subclass. A subclass is equivalent to a trivial, box-function class. 
The right inline figure at the bottom shows 3 such box functions representing the classical 3-class example (red, blue and their union).The formulation of \citet{qin2017wasserstein} supports only such classes. It is still as expressive as ours (2-class) but does not provide means to easily specify trade-offs between many objectives as it is not designed to scale to large number of objectives. Finally, we do not need to explicitly specify a target distribution for the union, which ends up being optimized toward a barycenter of the two targets.  \Cref{fig:Formulation}b shows one such example point set where the target distributions are only defined for the red and blue points. The union is optimized towards their barycenter following~\cref{eq:MultiClassBarycenterDiscrete}.

\section{Stochastic gradient-descent optimization}
\label{sec:Stochastic_gradient-descent_optimization}

In its most general form, the multi-class barycenter problem~\eqref{eq:MultiClassBarycenter} is continuous. For a finite number of optimization points, the effective number of subclasses within any class is finite too. However, the use of continuous class functions makes this number very large, far beyond the few objectives that state-of-the-art multi-class methods~\citep{wei2010multiclass,jiang2015bluenoise,qin2017wasserstein} can scale to, in terms of both memory and computation time. This is because these iterative methods optimize for all objectives at every step.

Taking cues from stochastic gradient-descent methods~\citep{Bottou:GradientDescent}, our approach is to optimize one objective at each of many iterations. Such optimization trivially scales to arbitrarily many objectives, although with potentially reduced convergence speed. Another advantage of this approach is that memory consumption does not directly depend on the objective count.

\Paragraph{Sliced multi-class barycenter}

Our multi-class barycenter formulation~\eqref{eq:MultiClassBarycenter} computing optimal transport plans and minimizing Wasserstein distances, which can be very costly. For practical efficiency, we turn to sliced optimal transport, replacing the Wasserstein distance $W_p$ by its sliced approximation $SW_p$~\eqref{eq:SlicedWassersteinDistance}. This adds another dimension to the integral in~\cref{eq:MultiClassBarycenter}, over the projection axis $\theta$:
\begin{tcolorbox}[ams equation,after=,]
    \label{eq:SlicedMultiClassBarycenter}
    \PointSetExt \,=\, \argmin_{\PointSetExt} \int_0^1 \!\!\! \int_0^1 \!\!\! \int_\SphereD \!\!\! W_p^p\big(\,\PointSetExt^\theta_{\ClassFunc_\ClassID>\Slice},\ProjectedTargetMeasure_{\ClassID,\ClassFunc_\ClassID>\Slice} \big) \,\dif \theta \dif \Slice \dif \ClassID.
\end{tcolorbox}
Since the sliced Wasserstein distance~\eqref{eq:SlicedWassersteinDistance} bounds the regular Wasserstein distance~\eqref{eq:WassersteinDistance}~\cite{bonnotte2013unidimensional}, the resulting optimization problem~\eqref{eq:SlicedMultiClassBarycenter} is an upper bound for the one in \cref{eq:MultiClassBarycenter}. During optimization, we use this~\cref{eq:SlicedMultiClassBarycenter} as our cost function which involves \emph{filtering} point set for each slice $\theta$. For brevity, we refer to this as filtered sliced optimal transport (FSOT) in the rest of the paper.

\Paragraph{Iterative minimization}

The 1D subclass Wasserstein distance in \cref{eq:SlicedMultiClassBarycenter} has a known solution whose derivative \wrt an optimization point $\Point_i$ we derive in \cref{sec:WasersteinDerivative}. The derivative of the entire barycenter is then a nested integral of such 1D derivatives. This enables an iterative stochastic minimization scheme which performs repeated 1D gradient-optimization steps by randomly sampling the multi-dimensional integral in \cref{eq:SlicedMultiClassBarycenter}.

\begin{figure}[t]
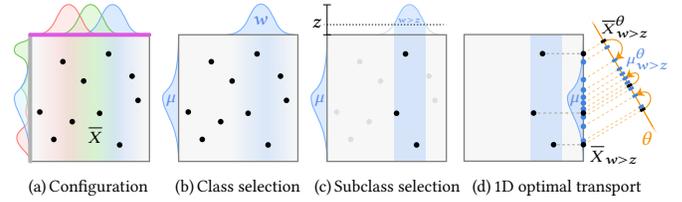

    \begin{overpic}[abs,unit=0.25mm]{figures/slicy-dicy}
        \fontsize{6.5}{7}\selectfont
        \put(8,3){(a)\,Configuration}
        \put(86,3){(b)\,Class selection}
        \put(160,3){(c)\,Subclass selection}
        \put(244,3){(d)\,1D optimal transport}
        \footnotesize
        \put(40,29){\scriptsize $\PointSetExt$}
        \put(127,92){\fontsize{7}{7}\selectfont\color{BlueClassColor} $\ClassFunc$}
        \put(81.3,50.4){\fontsize{6}{7}\selectfont\color{BlueClassColor} $\TargetMeasure$}
        \put(160.3,50.4){\fontsize{6}{7}\selectfont\color{BlueClassColor} $\TargetMeasure$}
        \put(296,50.4){\fontsize{6}{7}\selectfont\color{BlueClassColor} $\TargetMeasure$}
        \put(159,90){\fontsize{7}{7}\selectfont $\Slice$}
        \put(204.2,93){\fontsize{3.5}{7}\selectfont\color{BlueClassColor} $\ClassFunc \! > \! \Slice$}
        \put(334,28){\color{ProjectionColor} $\theta$}
        \put(307,19){\scriptsize $\PointSetExt_{\ClassFunc>z}$}
        \put(313,87){\fontsize{5}{7}\selectfont \rotatebox{0}{$\PointSetExt_{\ClassFunc>z}^\theta$}}
        \put(326,71){\fontsize{5}{7}\selectfont\color{BlueClassColor} $\TargetMeasure_{\ClassFunc>z}^\theta$}
    \end{overpic}
    \vspace{-4.5mm}
    \caption{
        One step of our stochastic gradient-descent minimization of the sliced multi-class barycenter~\eqref{eq:SlicedMultiClassBarycenter}. Given the optimization parameters and an extended point set~(a), we first select a class~(b) and then randomly threshold) the class function to sample a subclass~(c). Finally, we project the filtered set and the class' target distribution onto a sampled axis and perform one step of 1D gradient-descent optimization~(d). We handle arbitrary target distributions by point-sampling them before projection.
    }
    \label{fig:StochasticMinimization}
\end{figure}

\Cref{fig:StochasticMinimization} illustrates one step of our optimization procedure. Given an extended point set \PointSetExt and a class configuration, we first select a class $(\ClassFunc,\TargetMeasure)$, then threshold its function $\ClassFunc$ with a random value $\Slice$ to choose a subclass that isolates a fraction $\PointSetExt_{\ClassFunc > \Slice}$ of the points. Finally, we sample an axis $\theta$ and perform one step of gradient-descent optimization on the 1D Wasserstein distance between the projected points $\PointSetExt_{\ClassFunc > \Slice}^\theta$ and the projected (scaled) target distribution $\TargetMeasure_{\ClassFunc > \Slice}^\theta$ along the axis.

We repeat this entire process multiple times to obtain many offset vectors for every point. \Cref{sec:GradientEstimationPointOffsets} describes the computation of these offsets which are balanced across subclasses with varying sizes.
We average these offsets, update the point's position, and begin a new iteration on the result. This is similar to the method of \citet{paulin2020sliced} but simultaneously considering multiple optimization targets. Another difference is that we consider arbitrary target distributions on a general Euclidean domain $\Domain$.

Since 1D projections of target distributions cannot always be analytically represented, we point-sample them at a rate 3--5$\times$ higher than the \PointCount points being optimized. The optimization still solves a balanced (\ie discrete one-to-one) optimal-transport problem. This is done by binning the target points across \PointCount adaptive bins that follow the target distribution. Points within each bin are then averaged. \Cref{sec:GradientEstimationPointOffsets} provides more details.

\Paragraph{Offset correction}
\label{sec:OffsetCorrection}

Projecting a target distribution $\TargetMeasure$ along an arbitrary axis generally yields a different, non-uniform distribution $\ProjectedTargetMeasure$ for each axis, even when the target is uniform~\citep{paulin2020sliced}. Stochastic gradient descent on such different distributions can produce point offsets that are highly anisotropic in the (full-dimensional) optimization domain. The anisotropy is aligned with the density/domain boundaries and is susceptible to causing point alignments, as seen in \cref{fig:pointsets_comparison}c. We avoid this problem by scaling the gradient of each projected point by the projected target density at that point. We estimate this density using the projected target samples. For single-class sampling, this gradient correction is the major change between our method and \citet{paulin2020sliced} which leads to the quality improvement shown in~\cref{fig:pointsets_comparison}f. More details on the computation of this factor can be found in~\cref{sec:GradientEstimationPointOffsets}.

\Paragraph{Discussion}

The sliced Wasserstein distance is only an approximation to the regular distance, and can yield suboptimal barycenters~\citep{Bonneel2013SlicedWB}. However, in our experience it is a practical option for optimizing many points for many targets and produces satisfactory results even with highly non-uniform target distributions. Other approaches such as entropic regularization~\citep{cuturi2013sinkhorn}, stochastic barycenters~\citep{claici2018stochastic} or neural solvers~\citep{korotin2022neural} can also be employed but we leave that for future work. Another consequence of using sliced optimal transport is that it increases the effective number of objectives, by adding an extra (spherical) dimension to the barycentric integral~\eqref{eq:SlicedMultiClassBarycenter}. Thankfully, the individual ``sliced'' 1D objectives are simple, and stochastic optimization scales to the added complexity.

\section{Perceptual error optimization}
\label{sec:ErrorOptimization}

In this section we describe how to use our multi-class framework for perceptual optimization of image error in Monte-Carlo rendering.

In rendering, the value of every pixel is a light-transport integral. In practice pixel integrals are estimated via point sampling, and the resulting error manifests itself as image noise. Research efforts in sampling have traditionally focused on reducing the \emph{magnitude} of the error, \ie the accuracy of individual pixel estimates. Recently, it has been recognized that the \emph{distribution} of this error over the image plays an important perceptual role, and that visual fidelity can be drastically improved when this distribution is isotropic and high-frequency~\citep{georgiev2016blue}. Achieving such a blue-noise distribution requires carefully coordinating the samples \emph{across pixels}. We show that this problem can be cast as a multi-class optimization. We derive an image-error bound which can be minimized using our multi-class barycenter~\eqref{eq:MultiClassBarycenter}. The resulting formulation provides a principled way to minimize error in Monte Carlo rendering \wrt given perceptual and/or pixel-reconstruction kernels.

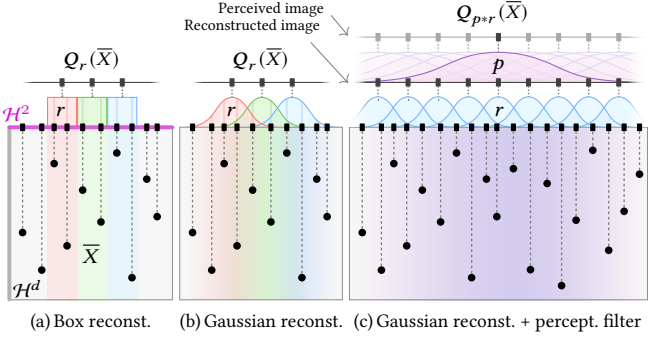
\begin{figure}[t!]
    \centering

\definecolor{OutlineColor1}{gray}{0.23}
\definecolor{OutlineColor2}{gray}{0.23}

\newcommand{\OneImage}[1]{
    \begin{scope}
        \clip (0.0,0.0) -- (9.5,0.0) -- (9.5,4.5) -- (0.0,4.5) -- cycle;
        \path[fill overzoom image=figures/#1] (0,0) rectangle (9.5cm,4.5cm);
    \end{scope}
}

\newcommand{\TestVeachImage}[2]{
    \begin{scope}
        \clip (0.0,4.5) -- (9.5,4.5) -- (9.5,7) -- (0.0,7) -- cycle;
        \path[fill overzoom image=figures/veach_mis/#1.jpg] (0,0) rectangle (9.5cm,7cm);
    \end{scope}
    \begin{scope}
        \clip (0.0,0.0) -- (9.5,0.0) -- (9.5,4.5) -- (0.0,4.5) -- cycle;
        \path[fill overzoom image=figures/veach_mis/#2.jpg] (0,0) rectangle (9.5cm,7cm);
    \end{scope}

    \begin{scope}
       \draw[draw=white,ultra thick] (0.0,4.5) -- (9.5,4.5);
       \draw[draw=white,ultra thick] (0.0,2.1) -- (9.5,2.1);
       \draw[draw=white,ultra thick] (2.5,0) -- (2.5,4.5);
      \draw[draw=white,ultra thick] (5,0) -- (5,2.1);
    \end{scope}
}

\newcommand{\VeachCrop}[5]{
    \begin{scope}
        \clip (#1,#2) -- (#3,#2) -- (#3,#4) -- (#1,#4) -- cycle;
        \path[fill overzoom image=figures/veach_mis/#5.jpg] (0,0) rectangle (9.5cm,7cm);
    \end{scope}
}

\newcommand{\VeachImage}[6]{
    \begin{scope}
        \clip (0.0,0.0) -- (2.5,0.0) -- (2.5,7) -- (0.0,7) -- cycle;
        \path[fill overzoom image=figures/veach_mis/#2.jpg] (0,0) rectangle (9.5cm,7cm);
    \end{scope}
    \begin{scope}
        \clip (2.5,2.1) -- (9.5,2.1) -- (9.5,7) -- (2.5,7) -- cycle;
        \path[fill overzoom image=figures/veach_mis/#3.jpg] (0,0) rectangle (9.5cm,7cm);
    \end{scope}
    \begin{scope}
        \clip (2.5,0.0) -- (5,0.0) -- (5,2.1) -- (2.5,2.1) -- cycle;
        \path[fill overzoom image=figures/veach_mis/#5.jpg] (0,0) rectangle (9.5cm,7cm);
    \end{scope}
    \begin{scope}
        \clip (5.0,0.0) -- (9.5,0.0) -- (9.5,2.1) -- (5.0,2.1) -- cycle;
        \path[fill overzoom image=figures/veach_mis/#6.jpg] (0,0) rectangle (9.5cm,7cm);
    \end{scope}

    \begin{scope}
    \end{scope}
}

\small
\hspace*{-2.3mm}
\begin{tabular}{c@{}}
\begin{tikzpicture}[scale=0.9]
    \OneImage{reconstruction-problem.ai}
    \begin{scope}
        \filldraw[black,ultra thick] (1.25, -0.2) circle (0pt) node[anchor=base,rotate=0]{\fontsize{7.3}{6}\selectfont (a)\,Box reconst.};
        \filldraw[black,ultra thick] (3.77, -0.2) circle (0pt) node[anchor=base,rotate=0]{\fontsize{7.3}{6}\selectfont (b)\,Gaussian reconst.};
        \filldraw[black,ultra thick] (7.26, -0.2) circle (0pt) node[anchor=base,rotate=0]{\fontsize{7.3}{6}\selectfont (c)\,Gaussian reconst.\ + percept.\ filter};
    \end{scope}
    
    \begin{scope}
        \filldraw[black,ultra thick] (1.0, 0.6) circle (0pt) node[anchor=south west,rotate=0]{$\PointSetExt$};
    \end{scope}
    
    \begin{scope}[overlay]
        \filldraw[black,ultra thick] (-0.2, 2.7) circle (0pt) node[anchor=south west,rotate=0]{\footnotesize\color{ExtSpaceColor}$\Hypercube^2$};
        \filldraw[black,ultra thick] (-0.05, 0.1) circle (0pt) node[anchor=south west,rotate=0]{\footnotesize\color{DomainColor}$\Hypercube^d$};
        \filldraw[black,ultra thick] (0.7, 3.45) circle (0pt) node[anchor=south west,rotate=0]{\footnotesize $\EstimatedImage_\ReconstFilter(\PointSetExt)$};
        \filldraw[black,ultra thick] (3.2, 3.45) circle (0pt) node[anchor=south west,rotate=0]{\footnotesize $\EstimatedImage_\ReconstFilter(\PointSetExt)$};
        \filldraw[black,ultra thick] (6.5, 4.1) circle (0pt) node[anchor=south west,rotate=0]{\footnotesize $\EstimatedImage_{\PerceptFilter * \ReconstFilter}(\PointSetExt)$};
    \end{scope}
    
    \begin{scope}
        \filldraw[black,ultra thick] (0.6, 2.78) circle (0pt) node[anchor=south west,rotate=0]{$\ReconstFilter$};
        \filldraw[black,ultra thick] (3.12, 2.78) circle (0pt) node[anchor=south west,rotate=0]{$\ReconstFilter$};
        \filldraw[black,ultra thick] (7.08, 2.78) circle (0pt) node[anchor=south west,rotate=0]{$\ReconstFilter$};
        \filldraw[black,ultra thick] (7.05, 3.4) circle (0pt) node[anchor=south west,rotate=0]{$\PerceptFilter$};
    \end{scope}
        
    \begin{scope}
        \filldraw[black,ultra thick] (3, 4.25) circle (0pt) node[anchor=south west,rotate=0]{\scriptsize Perceived image};
        \filldraw[black,ultra thick] (2.5, 4.0) circle (0pt) node[anchor=south west,rotate=0]{\scriptsize Reconstructed image};
        
        \draw[gray, ->] (4.5,4.0) -- (5.05,3.4);
        \draw[gray, ->] (4.8,4.4) -- (5.05,4.1);
    \end{scope}
    
\end{tikzpicture}
\end{tabular}
    \vspace{-2.5mm}
    \caption{
        Illustration of image synthesis where the grey box represents the sampling space, i.e.\ the unit hypercube $\Hypercube^{2+d}$; the horizontal axis represents the image subspace where reconstruction from the samples $\PointSetExt$ is performed.
        (a)~When using a box reconstruction kernel $\ReconstFilter$, the sample sets estimating different pixels are disjoint.
        (b)~A Gaussian kernel introduces overlaps, making each sample contribute to multiple pixel estimates.
        (c)~The human visual system applies additional filtering on the reconstructed image with a generally wider kernel~$\PerceptFilter$. The convolution $\PerceptFilter * \ReconstFilter$ acts as an effective reconstruction kernel for the \emph{perceived} image, and introduces even more overlaps.
    }
    \label{fig:reconstruction_problem_illustration}
\end{figure}

\subsection{Problem statement}

Given a point set $\PointSet = \{ \Point_i \}_{i=1}^\PointCount$, the value $\ReferencePixel_\ReconstFilter$ of an image pixel is estimated by point-sampling its associated integral:
\begin{equation}
    \EstimatedPixel_\ReconstFilter(\PointSet) = \frac{1}{\PointCount} \! \sum_{i=1}^\PointCount \! \ReconstFilter(\Point_i) f(\Point_i) \,\approx\, \ReferencePixel_\ReconstFilter 
     = \!\! \int_{\Hypercube^{2+\Dim}} \!\!\!\!\!\!\! \ReconstFilter(\Point) f(\Point) \,\dif \LebesgueMeasure(\Point) = \langle \ReconstFilter, f \rangle.
\end{equation}
Here \ReconstFilter is a pixel-reconstruction kernel, $f(\Point)$ is the illumination carried by a light-transport path corresponding to the point \Point in the unit hypercube $\Hypercube^{2+\Dim}$ with Lebesgue measure $\LebesgueMeasure$. The first two dimensions are image space (where $\ReconstFilter$ operates), and $\Dim$ is the path-space dimension. The variance of an estimate $\EstimatedPixel_\ReconstFilter(\PointSet)$ is reduced when the samples in \PointSet within the kernel support are well-stratified.

When using box-kernel reconstruction~(\cref{fig:reconstruction_problem_illustration}a), every sample falls within the kernel of a single pixel, which allows stratifying the samples independently per pixel. Non-box kernels, \eg Gaussians, generally overlap in image space, making each sample contribute to the estimates of several pixels (\cref{fig:reconstruction_problem_illustration}b). This case calls for coordinating the stratification of samples across pixels.

Moreover, our eyes do not perceive individual pixels but rather process the image as a whole. One type of processing that occurs is pre-filtering the input visual signal to avoid aliasing. That is, we perceive a version of the image that is blurred by an amount dependent on the observing distance. This filtering can be modeled as a discrete convolution of the ($\ReconstFilter$-reconstructed) pixels with a perceptual filter $\PerceptFilter$ ~\citep{Gonzlez2006AlphaSH,NasanenVisibility} that can be well approximated by a Gaussian~\citep{Pappas1999}. Every pixel in the \emph{perceived ground-truth image} thus takes the form 
$
\PerceptFilter * \ReferencePixel_\ReconstFilter
    \,=\, \PerceptFilter * \langle \ReconstFilter, f \rangle
    \,=\, \langle \PerceptFilter \hspace{-0.3mm} * \hspace{-0.3mm} \ReconstFilter, f \rangle
    \,=\, \ReferencePixel_{\PerceptFilter * \ReconstFilter}
$.
Analogously, pixels in the \emph{perceived estimated image} can be written as $\EstimatedPixel_{\PerceptFilter * \ReconstFilter}(\PointSet)$, which we illustrate in \cref{fig:reconstruction_problem_illustration}c. That image can thus be computed by convolving the samples with a combined reconstruction kernel $\PerceptFilter * \ReconstFilter$ centered at every pixel. The difference between the two images can be viewed as a measure of perceptual error~\citep{chizhov2022perceptual}. We can then formulate our problem as minimizing reconstruction \wrt a given (combined) kernel by optimizing the distribution of the samples \PointSet.

Note that in reality pixel reconstruction is performed by the renderer---to compute pixel estimates, while perceptual filtering occurs in the human visual system upon perceiving these estimates.

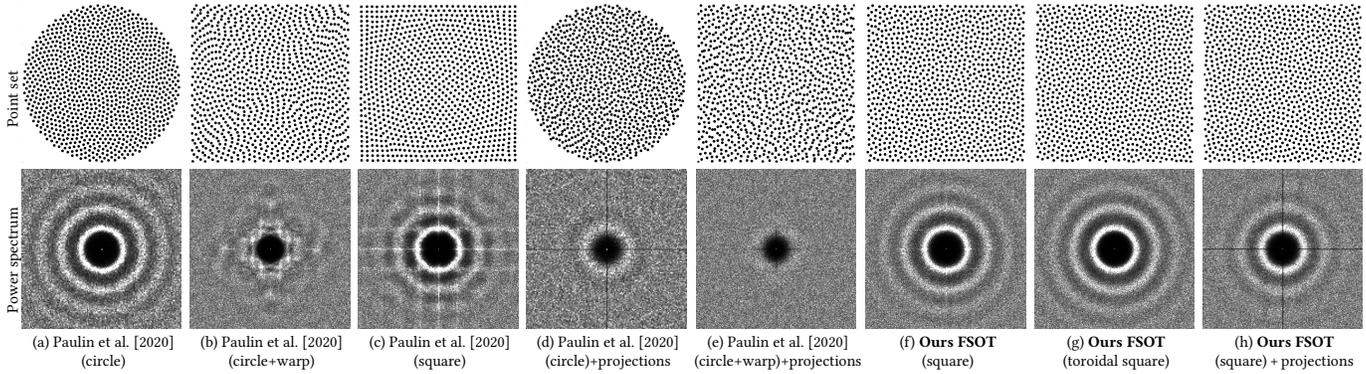
\begin{figure*}[t!]

\newcommand{\PlotSingleImage}[1]{
    \begin{tikzpicture}[scale=0.85]
        \begin{scope}
            \clip (0,0) -- (2.5,0) -- (2.5,2.5) -- (0,2.5) -- cycle;
            \path[fill overzoom image=figures/pointset_comparison/#1] (0,0) rectangle (2.5cm,2.5cm);
        \end{scope}
    \end{tikzpicture}
}

\scriptsize
\hspace*{-3.3mm}
\begin{tabular}{l@{}c@{\;}c@{\;}c@{\;}c@{\;}c@{\;}c@{\;}c@{\;}c@{}}
    \rotatebox{90}{\;\quad\quad Point set}
    &
    \PlotSingleImage{visu_sot_hyperball_n1024_c1_d2}
    &
    \PlotSingleImage{visu_sot_n1024_c1_d2}
    &
    \PlotSingleImage{visu_hypercube_sot_n1024_c1_d2}
    &
    \PlotSingleImage{visu_sot_proj_hyperball_n1024_c1_d2_r0}
    &
    \PlotSingleImage{visu_sot_proj_mapped_n1024_c1_d2_r0}
    &
    \PlotSingleImage{visu_uniformDir_n1024_c1_d2}
    &
    \PlotSingleImage{visu_uniformDir_toroidal_n1024_c1_d2}
    &
    \PlotSingleImage{visu_cross_n1024_c1_d2}
    \\
    \rotatebox{90}{\quad Power spectrum}
    &
    \PlotSingleImage{spectrum_sot_hyperball_n1024_c1_d2}
    &
    \PlotSingleImage{spectrum_sot_n1024_c1_d2}
    &
    \PlotSingleImage{spectrum_hypercube_sot_n1024_c1_d2}
    &
    \PlotSingleImage{spectrum_sot_proj_hyperball_n1024_c1_d2}
    &
    \PlotSingleImage{spectrum_sot_proj_mapped_n1024_c1_d2}
    &
    \PlotSingleImage{spectrum_uniformDir_n1024_c1_d2}
    &
    \PlotSingleImage{spectrum_uniformDir_toroidal_n1024_c1_d2}
    &
    \PlotSingleImage{spectrum_cross_n1024_c1_d2}
    \\
    &
    (a)~\citet{paulin2020sliced}  & (b)~\citet{paulin2020sliced}  & (c)~\citet{paulin2020sliced} & 
    (d)~\citet{paulin2020sliced}  & (e)~\citet{paulin2020sliced} & 
    (f)~\textbf{Ours FSOT} & 
    (g)~\textbf{Ours FSOT} & (h)~\textbf{Ours FSOT}
    \\
    & (circle) & (circle+warp)  & (square) &(circle)+projections & (circle+warp)+projections & (square) & (toroidal square) & (square)\,+\,projections 
\end{tabular}
     \vspace{-2.5mm}
     \caption{
        Comparison between different variants of our optimization and that of \citet{paulin2020sliced} which we build upon. All point sets are of size 1024 and the Fourier power spectra are averaged over 10 realizations. For our method we show realizations constructed with and without toroidality. \citeauthor{paulin2020sliced} optimize on the unit circle~(a) and then warp the resulting point set to the unit square~(b); they also show direct unit-square optimization~(c). Both their variants yield alignments that our method avoids (f-h), largely thanks to our offset correction (described in \cref{sec:OffsetCorrection}). One can also prioritize certain projections which can be beneficial for Monte-Carlo integration (see \cref{fig:MC_integration}); here we choose the $x$- or $y$-axis in 30\% of the optimization steps (d, e, h).
     }
     \label{fig:pointsets_comparison}
 \end{figure*}

\subsection{Multi-class image-error bound}
\label{subsec:MultiClassErrorBound}

\Cref{fig:reconstruction_problem_illustration} illustrates graphically how image-error minimization can be viewed as a multi-class optimization problem. Mapping the problem to the language of \cref{sec:Theory}, the optimization domain is the $d$-dimensional unit hypercube, $\Domain = \Hypercube^d$, and, notably, the class dimension is not the unit line (\eg as in \cref{fig:FormulationExample}b) but the unit square, $\ClassDim = \Hypercube^2$. The regular and extended point sets are identical, $\PointSetExt = \PointSet$. Every pixel has an associated reconstruction kernel and defines a distinct class, all sharing the Lebesgue measure \LebesgueMeasure as their (uniform) target distribution. Next we show that the barycenter between these classes provides a bound for the (perceptual) error of the image.

\Paragraph{Pixel-error bound}

The error of a pixel w.r.t.\ some given kernel $\ClassFunc$ is the difference between the estimated value and the ground truth: $\ErrorPixel_\ClassFunc(\PointSet) = \big| \EstimatedPixel_\ClassFunc(\PointSet) - \ReferencePixel_\ClassFunc \big|$. This becomes a \emph{perceptual error} when the kernel $\ClassFunc \coloneqq \PerceptFilter * \ReconstFilter$ incorporates perceptual filtering. \citet{paulin2020sliced} recently showed that optimal transport can provide a bound on the estimation error of pixel during Monte Carlo integration. In~\cref{sec:WeightedErrorBoundDerivation} we provide a simple proof for this bound which for a pixel in our setting reads $\ErrorPixel_\ClassFunc(\PointSetExt) \leq L_{\ClassFunc \cdot f} W(\PointSetExt, \LebesgueMeasure)$, which is the product of the Lipschitz constant of the integrand $\ClassFunc \!\cdot\! f$ and the 1-Wasserstein distance between the point set and the uniform distribution. Unfortunately, this bound is not immediately useful: it measures the deviation of the entire point set $\PointSetExt$ from uniformity and does not capture the strong effect of the narrow-support kernel $\ClassFunc$ on each pixel estimate. We instead desire a bound tailored to the estimation of weighted integrals of the form $\int \! w \! \cdot \! f$. We derive such a bound for the pixel error in \cref{sec:WeightedErrorBoundDerivation}:
\begin{align}
    \label{eq:PixelErrorBound}
    \ErrorPixel_\ClassFunc(\PointSetExt) \, \leq \,\, L_f \! \int_{\Reals} \!  W(\PointSetExt_{\ClassFunc>\Slice},\LebesgueMeasure_{\ClassFunc>\Slice}) \,\dif \Slice
        \,=\, L_f B_1(\PointSetExt, \ClassFunc, \LebesgueMeasure),
\end{align}
where $B_1$ is the minimization objective of the 1-Wasserstein subclass barycenter~\eqref{eq:SubclassBarycenter}. Note that the kernel $\ClassFunc$ has moved from the Lipschitz constant to the Wasserstein distance.

\Paragraph{Image-error bound}

Our end goal is to minimize the total image error. Applying the bound from \cref{eq:PixelErrorBound} to each of $\PixelCount$ pixels yields a bound for the $L_1$ error:
\begin{align}
    \label{eq:ImageErrorBound}
    \sum_{i=1}^\PixelCount \ErrorPixel_{\ClassFunc_i}(\PointSetExt)
        \,\leq \,\, L_f \sum_{i=1}^\PixelCount B_1(\PointSetExt, \ClassFunc_i, \LebesgueMeasure).
\end{align}
This bound is a product of the Lipschitz constant of $f$ and a (discrete) $\PixelCount$-class barycenter~\eqref{eq:MultiClassBarycenter}. It postulates that to reduce the image error, we need to increase the uniformity of all subsets of $\PointSetExt = \PointSet$ given by the $\Slice$-filtering of every kernel (\ie class function) $\ClassFunc_i$.

\Cref{eq:ImageErrorBound} is based on the 1-Wasserstein distance $W_1$, but in practice we use our $W_2$-based optimization scheme from \cref{sec:Stochastic_gradient-descent_optimization} to minimize a sliced variant of the bound. This works because $SW_1$ is bounded by $SW_2$. We provide a derivation of the $SW_1$ gradient in supplemental Section~S1. Note that we do not optimize the image-space dimensions of the points which are fixed and used for classification.

\section{Experiments}
\label{sec:experiments}

To demonstrate the utility of our multi-class framework, we show the results of several experiments from CPU (C++) and GPU (CUDA) implementations. The C++ implementation of our stochastic gradient-descent optimization is parallelizable across the projections within each iteration. The CUDA one parallelizes over different operations (projections, sorting, averaging). The different point sets presented below have been generated on an NVIDIA Quadro RTX 8000 and Intel\textsuperscript{\textregistered} Core\textsuperscript{\texttrademark} i9-8950HK CPU\,@\,2.90GHz. All rendering results have been generated using PBRT-v3~\cite{pharr2016physically}. The supplemental material includes an HTML viewer with more results.

\begin{figure*}[t!]
    \centering

\newcommand{\PlotSingleImage}[1]{
    \begin{tikzpicture}[scale=0.755]
        \begin{scope}
            \clip (0,0) -- (2.5,0) -- (2.5,2.5) -- (0,2.5) -- cycle;
            \path[fill overzoom image=figures/multi_class_comp/2-class/#1] (0,0) rectangle (2.5cm,2.5cm);
        \end{scope}
        \draw (0,0) -- (2.5,0) -- (2.5,2.5) -- (0,2.5) -- cycle;
    \end{tikzpicture}
}

\newcommand{\PlotRAPSLeftMost}[1]{
    \!\!
    \begin{tikzpicture}[scale=0.72]
        \begin{scope}
            \clip (0.02,0) -- (2.5,0) -- (2.5,1.25) -- (0,1.25) -- cycle;
            \path[fill overzoom image=figures/multi_class_comp/2-class/#1] (0,0) rectangle (2.5cm,1.25cm);
        \end{scope}
        \begin{scope}
        \scriptsize
        \filldraw[thick] (0.1, -0.25) circle (0pt) node[anchor=base,rotate=0] {0};
        \filldraw[thick] (1.1, -0.25) circle (0pt) node[anchor=base,rotate=0] {1};
        \filldraw[thick] (2.1, -0.25) circle (0pt) node[anchor=base,rotate=0] {2};
        \filldraw[thick] (2.0, 0.1) circle (0pt) node[anchor=base,rotate=0] {\tiny $\text{freq.}/\sqrt{\PointCount}$};
        \filldraw[thick] (0.12, 0.5) circle (0pt) node[anchor=base,rotate=0] {1};
        \end{scope}
    \end{tikzpicture}
}

\newcommand{\PlotRAPS}[1]{
    \!\!
    \begin{tikzpicture}[scale=0.72]
        \begin{scope}
            \clip (0.02,0) -- (2.5,0) -- (2.5,1.25) -- (0,1.25) -- cycle;
            \path[fill overzoom image=figures/multi_class_comp/2-class/#1] (0,0) rectangle (2.5cm,1.25cm);
        \end{scope}
        \begin{scope}
        \scriptsize
        \filldraw[thick] (0.1, -0.25) circle (0pt) node[anchor=base,rotate=0] {0};
        \filldraw[thick] (1.1, -0.25) circle (0pt) node[anchor=base,rotate=0] {1};
        \filldraw[thick] (2.1, -0.25) circle (0pt) node[anchor=base,rotate=0] {2};
        \filldraw[thick] (0.12, 0.5) circle (0pt) node[anchor=base,rotate=0] {1};
        \end{scope}
    \end{tikzpicture}
}

\small
\hspace*{-3.8mm}
\begin{tabular}{l@{}l@{}l@{}l@{\;\,}l@{}l@{}l@{\;\,}l@{}l@{}l@{}}
    &
    \multicolumn{3}{c}{Wasserstein blue-noise sampling~\citep{qin2017wasserstein}}& \multicolumn{3}{c}{\textbf{Ours FSOT}} & \multicolumn{3}{c}{\textbf{Ours FSOT} toroidal}
    \\[0.35mm]
    &
    \PlotSingleImage{visu_Wass_n1024_c2_d2_r0_class0.png}
    &
    \PlotSingleImage{visu_Wass_n1024_c2_d2_r0_class1.png}
    &
    \PlotSingleImage{visu_Wass_n1024_c2_d2_r0_full.png}
    &
    \PlotSingleImage{visu_non_toroidal_n1024_c2_d2_r9_class0.png}
    &
    \PlotSingleImage{visu_non_toroidal_n1024_c2_d2_r9_class1.png}
    &
    \PlotSingleImage{visu_non_toroidal_n1024_c2_d2_r9_full.png}
    &
    \PlotSingleImage{visu_n1024_c2_d2_r4_class0.png}
    &
    \PlotSingleImage{visu_n1024_c2_d2_r4_class1.png}
    &
    \PlotSingleImage{visu_n1024_c2_d2_r4_full.png}
    \\[-0.4mm]
    \rotatebox{90}{\scriptsize \quad Power spectrum}
    &
    \PlotSingleImage{RPdat_Wass_n1024_c2_d2_class0.png}
    &
    \PlotSingleImage{RPdat_Wass_n1024_c2_d2_class1.png}
    &
    \PlotSingleImage{RPdat_Wass_n1024_c2_d2.png}
    &
    \PlotSingleImage{RPdat_non_toroidal_n1024_c2_d2_class0.png}
    &
    \PlotSingleImage{RPdat_non_toroidal_n1024_c2_d2_class1.png}
    &
    \PlotSingleImage{RPdat_non_toroidal_n1024_c2_d2.png}
    &
    \PlotSingleImage{RPdat_n1024_c2_d2_class0.png}
    &
    \PlotSingleImage{RPdat_n1024_c2_d2_class1.png}
    &
    \PlotSingleImage{RPdat_n1024_c2_d2.png}
    \\
    \rotatebox{90}{\scriptsize \;\quad Radial avg.}
    &
    \PlotRAPSLeftMost{RAPS_Wass_n1024_c2_d2_class0.pdf}
    &
    \PlotRAPS{RAPS_Wass_n1024_c2_d2_class1.pdf}
    &
    \PlotRAPS{RAPS_Wass_n1024_c2_d2.pdf}
    &
    \PlotRAPS{RAPS_non_toroidal_n1024_c2_d2_class0.pdf}
    &
    \PlotRAPS{RAPS_non_toroidal_n1024_c2_d2_class1.pdf}
    &
    \PlotRAPS{RAPS_non_toroidal_n1024_c2_d2.pdf}
    &
    \PlotRAPS{RAPS_toroidal_n1024_c2_d2_class0.pdf}
    &
    \PlotRAPS{RAPS_toroidal_n1024_c2_d2_class1.pdf}
    &
    \PlotRAPS{RAPS_toroidal_n1024_c2_d2.pdf}
\end{tabular}
    \vspace{-2.5mm}
    \caption{
        3-class (red, blue, red \& blue) optimization of $2048$ points (top row), along with the corresponding expected power spectra (middle row) and their radial averages (bottom row). Our optimization (bounded and toroidal) achieves similar quality to that of \citet{qin2017wasserstein} (bounded); ours takes 38\,sec on GPU and theirs takes about 1 hour on CPU. The spectral anisotropy in the left two results is due to point alignments near the boundaries.
    }
    \label{fig:multi_class_2_colors}
\end{figure*}
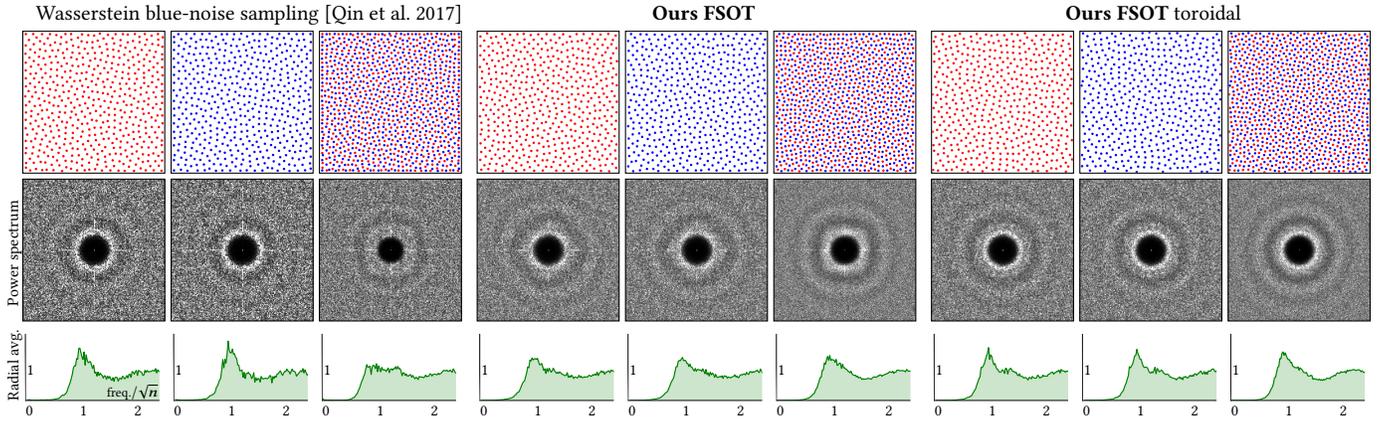

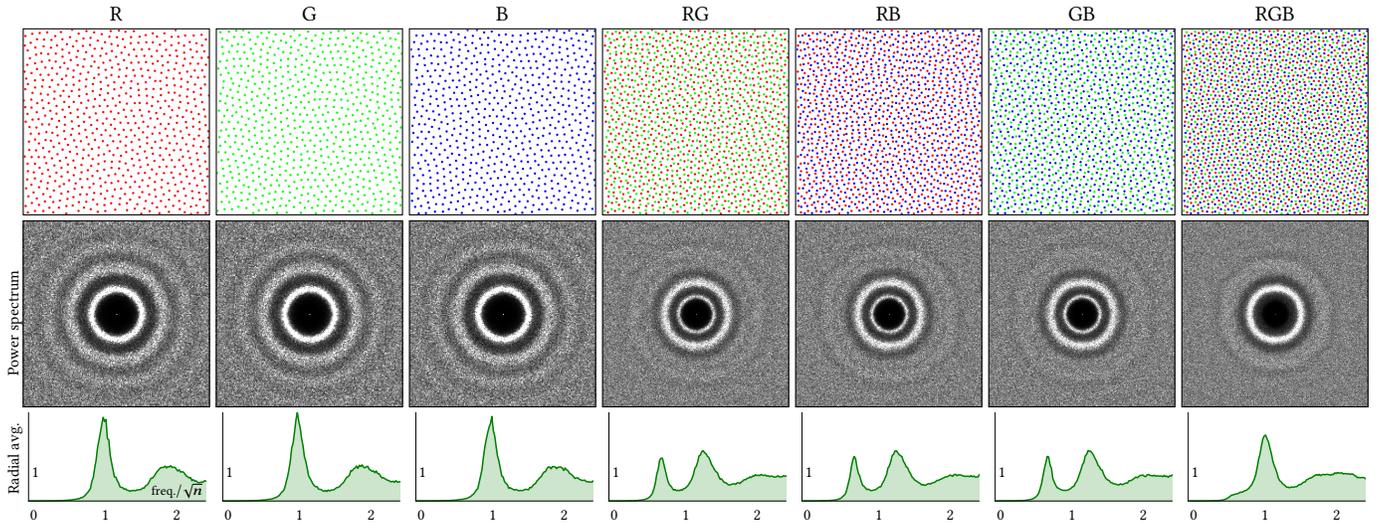
\begin{figure*}[t!]

\newcommand{\PlotSingleImage}[1]{%
    \begin{tikzpicture}[scale=0.99]
        \begin{scope}
            \clip (0,0) -- (2.5,0) -- (2.5,2.5) -- (0,2.5) -- cycle;
            \path[fill overzoom image=figures/multi_class_comp/#1] (0,0) rectangle (2.5cm,2.5cm);
        \end{scope}
        \draw (0,0) -- (2.5,0) -- (2.5,2.5) -- (0,2.5) -- cycle;
    \end{tikzpicture}%
}
\newcommand{\PlotRAPSLeftMost}[1]{%
    \begin{tikzpicture}[scale=0.95]
        \begin{scope}
            \clip (0,0) -- (2.5,0) -- (2.5,1.25) -- (0,1.25) -- cycle;
            \path[fill overzoom image=figures/multi_class_comp/#1] (0,0) rectangle (2.5cm,1.25cm);
        \end{scope}
        \begin{scope}
        \scriptsize
        \filldraw[thick] (0.1, -0.25) circle (0pt) node[anchor=base,rotate=0] {0};
        \filldraw[thick] (1.1, -0.25) circle (0pt) node[anchor=base,rotate=0] {1};
        \filldraw[thick] (2.1, -0.25) circle (0pt) node[anchor=base,rotate=0] {2};
        \filldraw[thick] (2.1, 0.1) circle (0pt) node[anchor=base,rotate=0] {\tiny $\text{freq.}/\sqrt{\PointCount}$};
        \filldraw[thick] (0.12, 0.35) circle (0pt) node[anchor=base,rotate=0] {1};
        \end{scope}
    \end{tikzpicture}%
}

\newcommand{\PlotRAPS}[1]{%
    \begin{tikzpicture}[scale=0.95]
        \begin{scope}
            \clip (0,0) -- (2.5,0) -- (2.5,1.25) -- (0,1.25) -- cycle;
            \path[fill overzoom image=figures/multi_class_comp/#1] (0,0) rectangle (2.5cm,1.25cm);
        \end{scope}
        \begin{scope}
        \scriptsize
        \filldraw[thick] (0.1, -0.25) circle (0pt) node[anchor=base,rotate=0] {0};
        \filldraw[thick] (1.1, -0.25) circle (0pt) node[anchor=base,rotate=0] {1};
        \filldraw[thick] (2.1, -0.25) circle (0pt) node[anchor=base,rotate=0] {2};
        \filldraw[thick] (0.12, 0.35) circle (0pt) node[anchor=base,rotate=0] {1};
        \end{scope}
    \end{tikzpicture}%
}

\small
\hspace*{-3.8mm}
\begin{tabular}{l@{}c@{\;}c@{\;}c@{\;}c@{\;}c@{\;}c@{\;}c@{}}
& R  & G  & B & RG & RB & GB & RGB 
\\
&
\PlotSingleImage{3-class/visu_uniformDir_n2048_c3_d2_r0_iter8000_class0_class0}
&
\PlotSingleImage{3-class/visu_uniformDir_n2048_c3_d2_r0_iter8000_class2_class2}
&
\PlotSingleImage{3-class/visu_uniformDir_n2048_c3_d2_r0_iter8000_class1_class1}
&
\PlotSingleImage{3-class/visu_uniformDir_n2048_c3_d2_r0_iter8000_class02_full}
&
\PlotSingleImage{3-class/visu_uniformDir_n2048_c3_d2_r0_iter8000_class01_full}
&
\PlotSingleImage{3-class/visu_uniformDir_n2048_c3_d2_r0_iter8000_class23_full}
&
\PlotSingleImage{3-class/visu_uniformDir_n2048_c3_d2_r0_iter8000_class2_full}
\\[-0.4mm]
\rotatebox{90}{\scriptsize \quad\quad Power spectrum}
&
\PlotSingleImage{3-class/RPdat_uniformDir_n2048_c3_d2_class0}
&
\PlotSingleImage{3-class/RPdat_uniformDir_n2048_c3_d2_class1}
&
\PlotSingleImage{3-class/RPdat_uniformDir_n2048_c3_d2_class2}
&
\PlotSingleImage{3-class/RPdat_uniformDir_n2048_c3_d2_class0_1}
&
\PlotSingleImage{3-class/RPdat_uniformDir_n2048_c3_d2_class0_2}
&
\PlotSingleImage{3-class/RPdat_uniformDir_n2048_c3_d2_class1_2}
&
\PlotSingleImage{3-class/RPdat_uniformDir_n2048_c3_d2}
\\[-0.3mm]
\rotatebox{90}{\scriptsize \quad\quad Radial avg.}
&
\PlotRAPSLeftMost{3-class/RAPS_n2048_c3_d2_class0}
&
\PlotRAPS{3-class/RAPS_n2048_c3_d2_class1}
&
\PlotRAPS{3-class/RAPS_n2048_c3_d2_class2}
&
\PlotRAPS{3-class/RAPS_n2048_c3_d2_class0_1}
&
\PlotRAPS{3-class/RAPS_n2048_c3_d2_class0_2}
&
\PlotRAPS{3-class/RAPS_n2048_c3_d2_class1_2}
&
\PlotRAPS{3-class/RAPS_n2048_c3_d2}
\end{tabular}
     \vspace{-2.5mm}
     \caption{
        Extending the problem in \cref{fig:multi_class_2_colors} to three colors (\ie 7 classes), using $2049$ points (683 points per color). Achieving uniform blue-noise quality across all classes is more difficult in this case due to higher, contention between the objectives.
     }
     \label{fig:multi_class_3_colors}
\end{figure*}

\begin{figure*}[t!]

\newcommand{\OneImage}[1]{%
    \begin{tikzpicture}[scale=0.659]
        \begin{scope}
            \clip (0,0)-- (0.0,8.5) -- (4.5,8.5) -- (4.5,0.0) -- cycle;
            \path[fill overzoom image=figures/#1] (0,0) rectangle (4.5,8.5);
        \end{scope}
    \end{tikzpicture}%
}

\small
\hspace*{-2.5mm}
\begin{tabular}{c@{\;}c@{\,}c@{\,}c@{\,}c@{\,}c@{}}
    \OneImage{stipling/beach/ref.jpeg}
    &
    \OneImage{stipling/beach/beach_cmyk.png}
    &
    \OneImage{stipling/beach/beach_k.png}
    &
    \OneImage{stipling/beach/beach_m.png}
    &
    \OneImage{stipling/beach/beach_cy.png}
    &
    \OneImage{stipling/beach/beach_cmy.png}
    \\[-0.2mm]
    Reference & All colors (CMYK) & Black (K) & Magenta (M) & Cyan + yellow (CY) & (CMY)
\end{tabular}
 
    \vspace{-2.5mm}
    \caption{
        CMYK color stippling involves optimizing 15 classes---four base colors and their various 2- and 3-color combinations, each targeting a different density. In this example we use 20,000 points and show five of these classes.
    }
    \label{fig:stippling2}
\end{figure*}
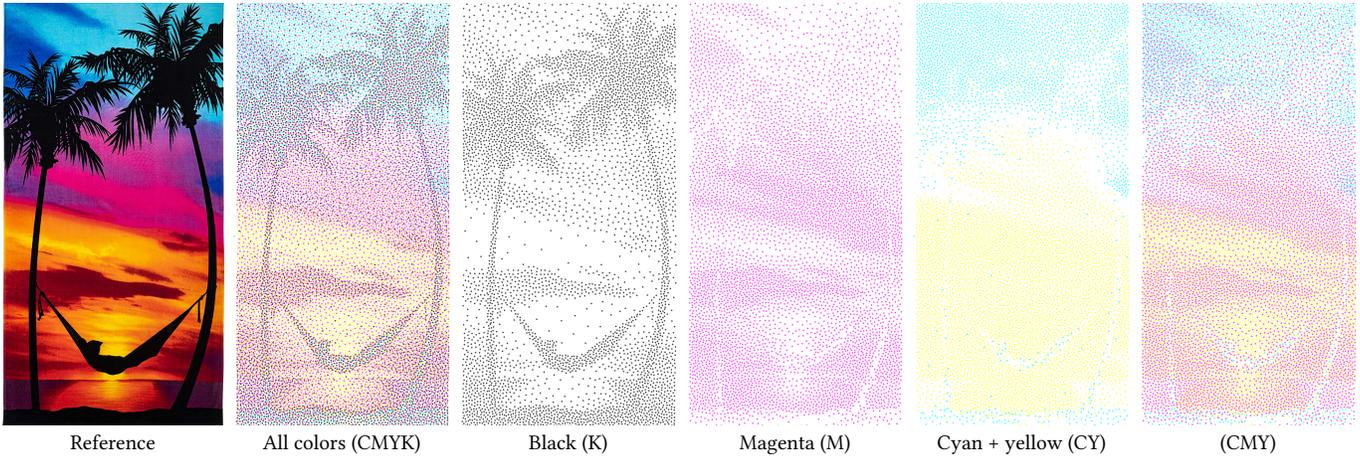

\begin{figure*}[t!]

\newcommand{\PlotSingleImage}[1]{%
        \begin{scope}
            \clip (0,0) -- (2.5,0) -- (2.5,2.5) -- (0,2.5) -- cycle;
            \path[fill overzoom image=figures/continous_class_extract/#1] (0,0) rectangle (2.5cm,2.5cm);
        \end{scope}
        \draw (0,0) -- (2.5,0) -- (2.5,2.5) -- (0,2.5) -- cycle;
}

\newcommand{\PlotRAPS}[1]{%
        \begin{scope}
            \clip (0,0) -- (2.5,0) -- (2.5,1.25) -- (0,1.25) -- cycle;
            \path[fill overzoom image=figures/continous_class_extract/#1] (0,0) rectangle (2.5cm,1.25cm);
        \end{scope}
        \begin{scope}
        \scriptsize
        \filldraw[thick] (0.1, -0.25) circle (0pt) node[anchor=base,rotate=0] {0};
        \filldraw[thick] (1.1, -0.25) circle (0pt) node[anchor=base,rotate=0] {1};
        \filldraw[thick] (2.1, -0.25) circle (0pt) node[anchor=base,rotate=0] {2};
        \filldraw[thick] (0.12, 0.5) circle (0pt) node[anchor=base,rotate=0] {1};
        \end{scope}
}
\newcommand{\PlotRAPSLeftMost}[1]{%
        \begin{scope}
            \clip (0,0) -- (2.5,0) -- (2.5,1.25) -- (0,1.25) -- cycle;
            \path[fill overzoom image=figures/continous_class_extract/#1] (0,0) rectangle (2.5cm,1.25cm);
        \end{scope}
        \begin{scope}
        \scriptsize
        \filldraw[thick] (0.1, -0.25) circle (0pt) node[anchor=base,rotate=0] {0};
        \filldraw[thick] (1.1, -0.25) circle (0pt) node[anchor=base,rotate=0] {1};
        \filldraw[thick] (2.1, -0.25) circle (0pt) node[anchor=base,rotate=0] {2};
        \filldraw[thick] (2.2, 0.1) circle (0pt) node[anchor=base,rotate=0] {\tiny $\text{freq.}/\sqrt{\PointCount}$};
        \filldraw[thick] (0.12, 0.5) circle (0pt) node[anchor=base,rotate=0] {1};
        \end{scope}
}

\newcommand{\TwoColumnFigure}[2]{%
    \begin{tabular}{c@{\;}c@{}}
        \hspace*{-2.5mm}
        \begin{tikzpicture}[scale=0.563]
            \PlotSingleImage{#1}
        \end{tikzpicture}
         & 
         \begin{tikzpicture}[scale=0.563]
            \PlotSingleImage{#2}
        \end{tikzpicture}
    \end{tabular}%
}

\small
\hspace*{-4mm}
\begin{tabular}{l@{}c@{\;}c@{\;}c@{\;}c@{\;}c@{\;}c@{}}
&
\textcolor{blue}{$80\%$} \textcolor{red}{$20\%$} &
\textcolor{blue}{$60\%$} \textcolor{red}{$40\%$} &
\textcolor{blue}{$50\%$} \textcolor{red}{$50\%$} &
\textcolor{blue}{$40\%$} \textcolor{red}{$60\%$} &
\textcolor{blue}{$30\%$} \textcolor{red}{$70\%$} &
\textcolor{blue}{$20\%$} \textcolor{red}{$80\%$} 
\\
&
\begin{tikzpicture}[scale=1.165]
    \PlotSingleImage{visu_progressive_n2048_c1_d2_r1_s4.png}
\end{tikzpicture}
&
\begin{tikzpicture}[scale=1.165]
    \PlotSingleImage{visu_progressive_n2048_c1_d2_r1_s6.png}
\end{tikzpicture}
&
\begin{tikzpicture}[scale=1.165]
    \PlotSingleImage{visu_progressive_n2048_c1_d2_r1_s8.png}
\end{tikzpicture}
&
\begin{tikzpicture}[scale=1.165]
    \PlotSingleImage{visu_progressive_n2048_c1_d2_r1_s10.png}
\end{tikzpicture}
&
\begin{tikzpicture}[scale=1.165]
    \PlotSingleImage{visu_progressive_n2048_c1_d2_r1_s12.png}
\end{tikzpicture}
&
\begin{tikzpicture}[scale=1.165]
    \PlotSingleImage{visu_progressive_n2048_c1_d2_r1_s16.png}
\end{tikzpicture}
\\[-0.4mm]
&
\TwoColumnFigure{visu_progressive_n2048_c1_d2_r1_subdiv_4_r}{visu_progressive_n2048_c1_d2_r1_subdiv_4_b}
&
\TwoColumnFigure{visu_progressive_n2048_c1_d2_r1_subdiv_6_r}{visu_progressive_n2048_c1_d2_r1_subdiv_6_b}
&
\TwoColumnFigure{visu_progressive_n2048_c1_d2_r1_subdiv_8_r}{visu_progressive_n2048_c1_d2_r1_subdiv_8_b}
&
\TwoColumnFigure{visu_progressive_n2048_c1_d2_r1_subdiv_10_r}{visu_progressive_n2048_c1_d2_r1_subdiv_10_b}
&
\TwoColumnFigure{visu_progressive_n2048_c1_d2_r1_subdiv_12_r}{visu_progressive_n2048_c1_d2_r1_subdiv_12_b}
&
\TwoColumnFigure{visu_progressive_n2048_c1_d2_r1_subdiv_16_r}{visu_progressive_n2048_c1_d2_r1_subdiv_16_b}
\\[-0.4mm]
&
\TwoColumnFigure{RPdat_coninous_d2_4_a}{RPdat_coninous_d2_4_b}
&
\TwoColumnFigure{RPdat_coninous_d2_6_a}{RPdat_coninous_d2_6_b}
&
\TwoColumnFigure{RPdat_coninous_d2_8_a}{RPdat_coninous_d2_8_b}
&
\TwoColumnFigure{RPdat_coninous_d2_10_a}{RPdat_coninous_d2_10_b}
&
\TwoColumnFigure{RPdat_coninous_d2_12_a}{RPdat_coninous_d2_12_b}
&
\TwoColumnFigure{RPdat_coninous_d2_16_a}{RPdat_coninous_d2_16_b}
\\[-0.4mm]
\rotatebox{90}{\scriptsize \quad\qquad Radial avg.}
&
\begin{tikzpicture}[scale=1.15]
    \PlotRAPSLeftMost{RAPS_continious_c1_d2_4}
\end{tikzpicture}
&
\begin{tikzpicture}[scale=1.15]
    \PlotRAPS{RAPS_continious_c1_d2_6}
\end{tikzpicture}
&
\begin{tikzpicture}[scale=1.15]
    \PlotRAPS{RAPS_continious_c1_d2_8}
\end{tikzpicture}
&
\begin{tikzpicture}[scale=1.15]
    \PlotRAPS{RAPS_continious_c1_d2_10}
\end{tikzpicture}
&
\begin{tikzpicture}[scale=1.15]
    \PlotRAPS{RAPS_continious_c1_d2_12}
\end{tikzpicture}
&
\begin{tikzpicture}[scale=1.15]
    \PlotRAPS{RAPS_continious_c1_d2_16}
\end{tikzpicture}
%
\end{tabular} 
     \vspace{-2.5mm}
     \caption{
        A set of 2048 points optimized according to a configuration with two linear-ramp-function classes. At every point index we can split the set into two subsets with good-quality distribution each, for an effective number of 4096 optimization targets (\ie subclasses). We show six such (color-coded) splits and the 2D Fourier power spectra of the two extracted subsets. The last row shows the radially averaged spectra of the subsets and the entire point set (in green).
     }
     \label{fig:continuous_class_extraction}
\end{figure*}
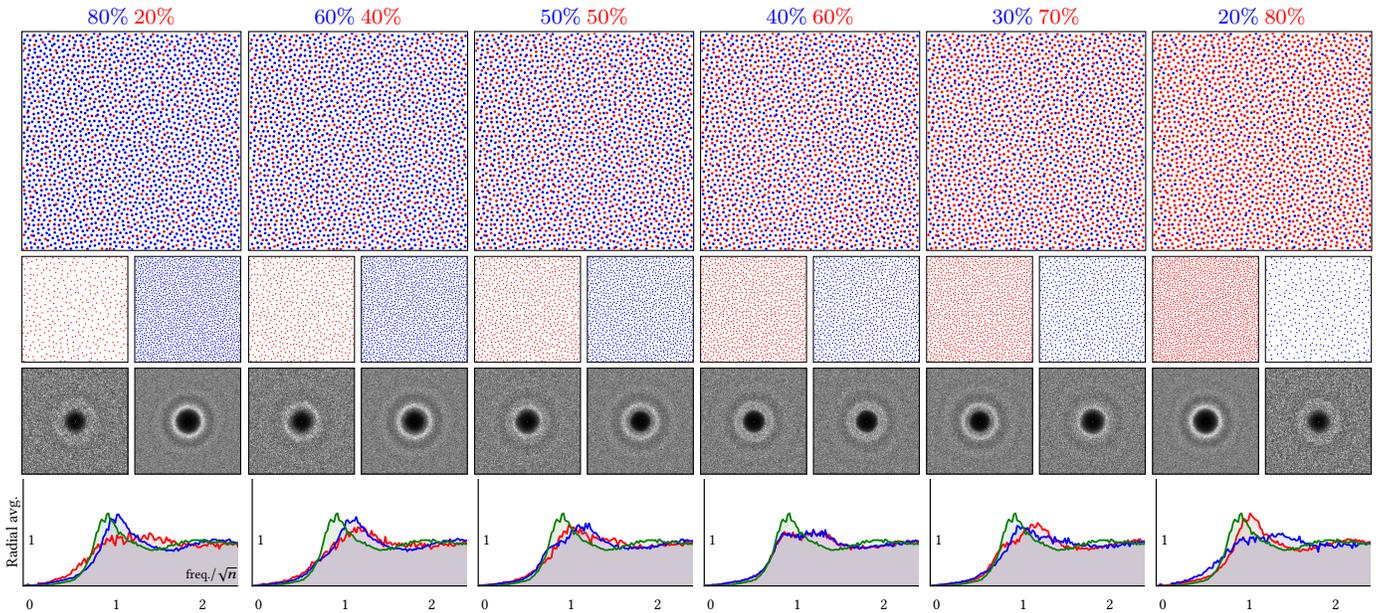

 \begin{figure}[t!]
    \centering

\newcommand{\TwoImage}[3]{
    \begin{scope}
        \clip (0,0) rectangle (3.75,6.5);
        \path[fill overzoom image=figures/#1] (0,0) rectangle (3.75,6.5);
    \end{scope}
    \begin{scope}
        \path[fill overzoom image=figures/#2] (0.0,4.0) rectangle ++(1.25cm,2.2cm);
    \end{scope}
    \begin{scope}
        \draw[line width=0.2mm,gray]  (0.0,4.0) rectangle ++(1.25cm,2.2cm);
    \end{scope}
    \begin{scope}
        \clip (0.1,0.1) -- (1.6,0.1) -- (1.6,1.6) -- (0.1,1.6) -- cycle;
        \path[fill overzoom image=figures//#3] (0.1,0.1) rectangle (2.1cm,1.9cm);
    \end{scope}
    \begin{scope}
        \draw[gray,thick] (0.1,0.1) -- (1.6,0.1) -- (1.6,1.6) -- (0.1,1.6) -- cycle;
    \end{scope}
}

\newcommand{\OneImage}[2]{
    \begin{scope}
        \clip (0,0)-- (0.0,6.5) -- (3.75,6.5) -- (3.75,0.0) -- cycle;
        \path[fill overzoom image=figures/#1] (0,0) rectangle (3.75,6.5);
    \end{scope}
    \begin{scope}
        \clip (0.1,0.1) -- (1.6,0.1) -- (1.6,1.6) -- (0.1,1.6) -- cycle;
        \path[fill overzoom image=figures//#2] (0.1,0.1) rectangle (2.1cm,1.9cm);
    \end{scope}
    \begin{scope}
        \draw[gray,thick] (0.1,0.1) -- (1.6,0.1) -- (1.6,1.6) -- (0.1,1.6) -- cycle;
    \end{scope}
}

\small
\begin{tabular}{c@{\;\;\;\;\;\;}c@{}}
    \begin{tikzpicture}[scale=1.0]
        \TwoImage{stipling/boody/boody_bnot.png}{stipling/boody/boody_ref.png}{stipling/boody/crop_bnot.png}
    \end{tikzpicture}
    &
    \begin{tikzpicture}[scale=1.0]
        \OneImage{stipling/boody/boody_ours.png}{stipling/boody/crop_ours.png}
    \end{tikzpicture}
    \\[-1mm]
    BNOT~\cite{deGoes2012blue} & \textbf{Ours FSOT}
\end{tabular}
     \vspace{-2mm}
     \caption{
        Comparison of monochrome image stippling using 15,000 points.
     }
     \label{fig:stippling_boody}
 \end{figure}
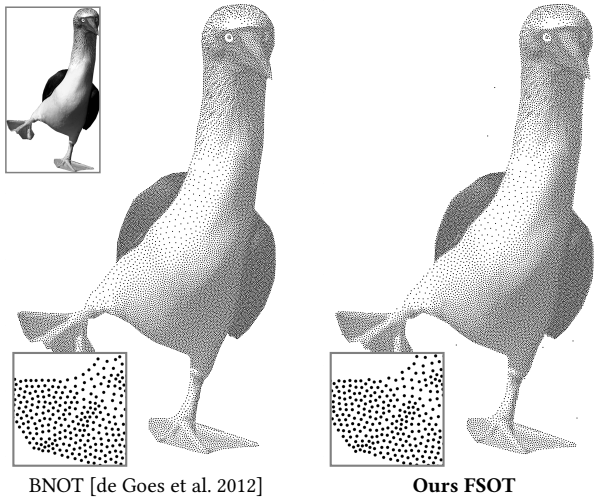

\subsection{Blue-noise sampling}

\Paragraph{Single-class blue noise}

\Cref{fig:pointsets_comparison} compares the blue-noise quality for a single-class point set and its power spectrum averaged over 10 realizations. \citet{paulin2020sliced} perform the optimization on the unit circle, achieving high quality (\cref{fig:pointsets_comparison}a) which, however, deteriorates after warping the points to the unit square (\cref{fig:pointsets_comparison}b); this is also reflected in the power spectrum. \citeauthor{paulin2020sliced} also show direct optimization on the unit square, which yields strong alignments along the domain boundaries (\cref{fig:pointsets_comparison}c). In contrast, our unit-square optimization produces a high-quality blue-noise distribution, without any alignments (see \cref{fig:pointsets_comparison}f). This quality improvement is mostly due to our offset correction (\cref{sec:OffsetCorrection}) which avoids alignments. Our optimization can also operate on a toroidal domain (\cref{fig:pointsets_comparison}g).

Prioritizing certain projection directions can be beneficial in Monte-Carlo integration as we will demonstrate below; \Cref{fig:pointsets_comparison}h shows an example where we choose the $x$- or $y$-aixs with 30\% probability, creating a cross in the power spectrum. While \citet{paulin2020sliced} can also prioritize these projections on the unit circle (\cref{fig:pointsets_comparison}d), the achieved quality is not maintained after mapping the points to the unit square (\cref{fig:pointsets_comparison}e).

\Paragraph{Multi-class sampling with uniform density}

In~\cref{fig:multi_class_2_colors}, we compare our method to that of \citet{qin2017wasserstein} on the two-color problem from \cref{fig:FormulationExample}. The spectra obtained by \citet{qin2017wasserstein} and our method without toroidality show some artefacts due to natural point alignments near the domain boundaries. Our method shows same quality for the single colors and slightly better for the complete set. In~\cref{fig:multi_class_3_colors}, we extend the problem to 3 colors, \ie 7 classes. The overall distribution quality is good for all classes. The spectral distributions of the three color pairs RG, RB, GB exhibit double peaks, which has also been observed by \citet[Fig.\,7]{qin2017wasserstein}. The reason for this double peak is that the improvement of these particular two-color classes has a strong impact on the other classes. Improving two-color classes would reduce the quality of the other classes too much.

\Paragraph{Color stippling}

\Cref{fig:stippling2} shows a CMYK image stippled with 20,000 points. The four individual colors and their various 2- and 3-color combinations each represent a class with a different target density, for a total of 15 classes. We show five of these classes. The combinations have weighted-average densities based on the respective energy of the channels. Unlike prior work~\citep{qin2017wasserstein}, our stochastic gradient-descent optimization scales to this many classes with a negligible memory footprint. The supplemental document shows another color-stippling result with 40,000 points.

\Cref{fig:stippling_boody} compares our stippling to that of \citet{deGoes2012blue} on a greyscale image using 15,000 points. Although our method is not tuned for single-class problems, we achieve competitive quality. The supplemental document includes a result with 100,000 points.

Our method also be used for animation stippling where consecutive frames share a fraction of the points. We include an example in the supplemental material.

\Paragraph{Continuous class extraction}

To demonstrate the scalability of our optimization to a large number of objectives, in \cref{fig:continuous_class_extraction} we consider a non-traditional multi-class problem. We define two classes
\WrapFigureTemplate{0.29}{3.8mm}{-2.8mm}{-3.7mm}{%
    \begin{overpic}[abs,unit=0.25mm,scale=1,grid=false]{figures/continuous-class}
    \end{overpic}%
}%
with linear-ramp functions on the index space of points, as illustrated in the inline figure. The target density is uniform. This construction allows us to split the optimized set at any point index, so that the subsets on the left and right of it (and their union) have good-quality distribution. We optimize 2048 points for an effective total number of 4096 targets (\ie subclasses). In the figure we include a few example splits; the corresponding subset power spectra show reasonable blue-noise quality considering the large number of optimization objectives. An animation showing the evolution of the visualization according to the choice of splitting index can be found in the supplemental material.

\begin{figure}[t!]
    \centering

\newcommand{\OneImage}[1]{
    \begin{scope}
        \clip (0,0)-- (0.0,4.5) -- (5.4,4.5) -- (5.4,0.0) -- cycle;
        \path[fill overzoom image=figures/#1] (0,0) rectangle (5.4,4.5);
        
    \end{scope}
}

\newcommand{\ProgressivePlotTwo}[2]{
    \begin{scope}
        \clip (0.0,0.0) -- (4.5,0.0) -- (4.5,4.5) -- (0.0,4.5) -- cycle;
        \path[fill overzoom image=figures//#1] (0,0) rectangle (4.5cm,4.5cm);
    \end{scope}
    \begin{scope}
        \clip (0.1,0.1) -- (1.6,0.1) -- (1.6,1.6) -- (0.1,1.6) -- cycle;
        \path[fill overzoom image=figures//#2] (0.1,0.1) rectangle (2.1cm,1.9cm);
    \end{scope}
    \begin{scope}
        \draw[black,thick] (0.1,0.1) -- (1.6,0.1) -- (1.6,1.6) -- (0.1,1.6) -- cycle;
    \end{scope}
}

\small
\hspace*{-2mm}
\begin{tabular}{c@{\;}c@{}}
\begin{tikzpicture}[scale=0.85]
    \ProgressivePlotTwo{object_placement/object_placement_vari_color.png}{object_placement/visu_continuous_class_obj_placement.png}
\end{tikzpicture}
&
\begin{tikzpicture}[scale=0.85]
    \OneImage{object_placement/object_placement_2.jpeg}
\end{tikzpicture}

\end{tabular} 
    \vspace{-2.5mm}
    \caption{
        Application of our continuous-class optimization to placement of objects with continuous variation in color (left) and size (right).
    }
    \label{fig:object_placement}
\end{figure}
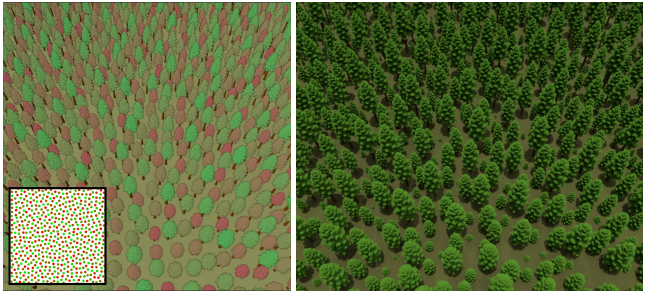

\Paragraph{Object placement}

Multi-class sampling can also be used to place objects in an environment. \Cref{fig:teaser} middle shows an example distribution of trees, each taking one of 7 colors. We also optimize for the union, for a total of 8 classes. Two other results are displayed in \cref{fig:object_placement}. The point set used (in the lower left corner) was produced using the optimization configuration from the continuous class extraction problem presented above. In the left image, the point color guides the tree color, and in the right image it guides the tree height.

\subsection{Monte-Carlo integration}

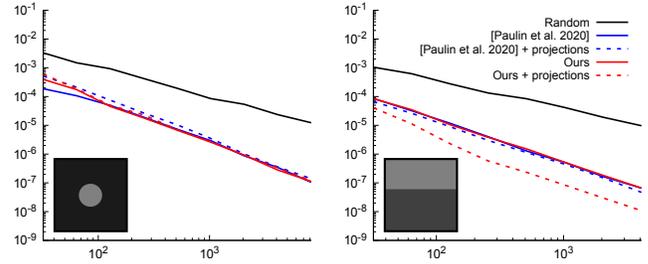
\begin{figure}[t!]
    \centering

\newcommand{\OneImage}[1]{
    \begin{scope}
        \clip (0,0)-- (0.0,4.5) -- (4.5,4.5) -- (4.5,0.0) -- cycle;
        \path[fill overzoom image=figures/#1] (0,0) rectangle (4.5,4.5);
        
    \end{scope}
}

\newcommand{\ProgressivePlotTwo}[2]{
    \begin{scope}
        \clip (0.0,0.0) -- (5.4,0.0) -- (5.4,4.5) -- (0.0,4.5) -- cycle;
        \path[fill overzoom image=figures//#1] (0,0) rectangle (5.4cm,4.5cm);
    \end{scope}
    \begin{scope}
        \clip (0.8,0.6) -- (2.0,0.6) -- (2.0,2.0) -- (0.8,2.0) -- cycle;
        \path[fill overzoom image=figures//#2] (0.8,0.6) rectangle (2.0cm,1.8cm);
    \end{scope}
    \begin{scope}
        \draw[black,thick] (0.8,0.6) -- (2.0,0.6) -- (2.0,1.8) -- (0.8,1.8) -- cycle;
    \end{scope}
}

\small
\hspace*{-2mm}
\begin{tabular}{c@{}c@{}}
\begin{tikzpicture}[scale=0.8]
    \ProgressivePlotTwo{MC_integration/graph_MCI_disc_v3.pdf}{MC_integration/disc.png}
\end{tikzpicture}
\begin{tikzpicture}[scale=0.8]
    \ProgressivePlotTwo{MC_integration/graph_MCI_alligned_step_v3.pdf}{MC_integration/alligned_step.png}
\end{tikzpicture}
\end{tabular} 
    \vspace{-2.5mm}
    \caption{
        Comparison of the Monte-Carlo variance convergence of our optimized point sets against those of \citet{paulin2020sliced}. We average variance over 10 realizations of each method and 40 variations of each function. Our axis-aligned projection prioritization is more effective than theirs.
    }
    \label{fig:MC_integration}
\end{figure}

We also evaluate our approach on Monte-Carlo integration. In \cref{fig:MC_integration} we analyze the convergence behavior of our optimized point sets against the method of \citet{paulin2020sliced} on two simple integrands. We generate two types of point sets using each method: one with axis-aligned 1D projections prioritized with 30\% probability (as in \cref{fig:pointsets_comparison}h) and the other without prioritization (as in \cref{fig:pointsets_comparison}g). For the isotropic integrand on the left the four variants give similar results. On the other hand, on the right integrand with axis-aligned variation, our projections yield lower integration error. Axis prioritization using \citeauthor{paulin2020sliced}'s method is ineffective since the post-optimization point warping to the unit square ruins the point-set properties.

\Paragraph{Progressive sampling}

\begin{figure}[t!]
    \centering
    \includegraphics{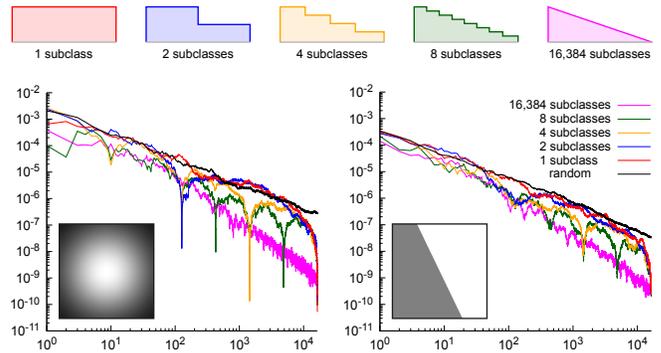} 
    \vspace{-2.5mm}
    \caption{
        Progressive point-set optimization using a single, staircase-function class. Increasing the number of steps (with equal lengths in power-of-2 scale) increases the number of prefix subsets to optimize; we show 5 examples. The graphs plot integration-error behavior with increasing number of points taken, up to 16,384, averaged over 30 integrand variations and 20 point-set realisations. We see that finer progressive granularity yields a larger number of error dips, but each is shorter. The fully progressive (pink) point set exhibits uniform error behavior.
    }
    \label{fig:progressive_sampling}
\end{figure}

Our multi-class formulation allows constructing progressive point patterns with controlled granularity. We can use a single, staircase-function class where the number of steps (\ie subclasses) dictates the number of prefix subsets (\ie progressive levels) to optimize for. A constant class function corresponds to optimizing only the full set of points for uniformity; in the other extreme of a linear-ramp class function every prefix of points is optimized. \Cref{fig:progressive_sampling} shows progressive error-convergence plots for 5 such variants using 16,384 points. The steps have equal lengths in power-of-2 scale. The 1-subclass red curve behaves almost like a random one for all sample counts except for the strong dip at the end. Only when all samples are used is the integration error low; in fact, this is the lowest error achieved by any point set in the plot. Increasing the number of subclasses increases the number of dips but also shortens each. This result clearly illustrates that finer progressive granularity comes at the cost of increased error due to the larger number of objectives the optimization needs to balance. In the extreme case of 16,384 subclasses, the point set is fully progressive and shows uniform error behavior.

\Paragraph{Rendering}

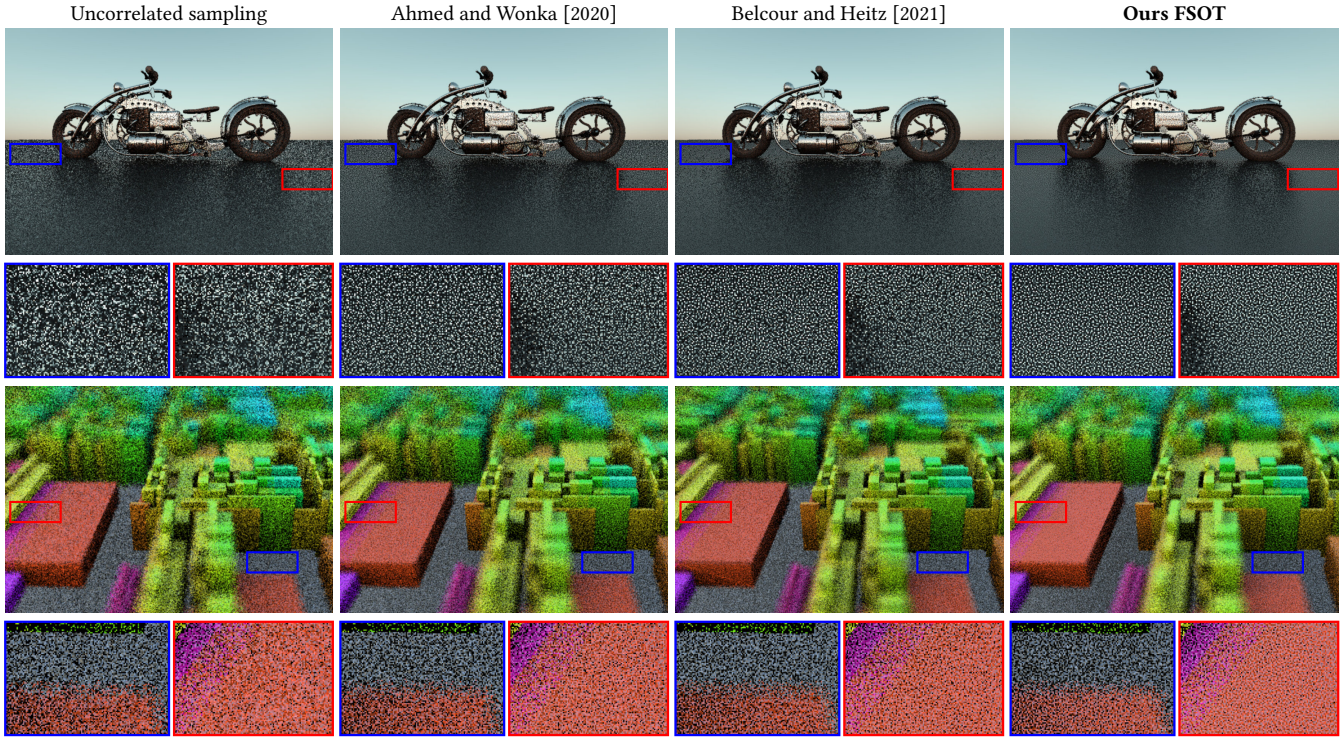
\begin{figure*}[t!]
    \centering

\newcommand{\HighDimExempleTemplate}[1]{
    \begin{scope}
        \clip (0,0) -- (6.5,0) -- (6.5,4.5) -- (0,4.5) -- cycle;
        \path[fill overzoom image=figures/#1.jpg] (0,0) rectangle (6.5cm,4.5cm);
        
    \end{scope}
}

\newcommand{\chopper}[1]{
    \begin{scope}
        \clip (0,0) -- (6.5,0) -- (6.5,4.5) -- (0,4.5) -- cycle;
        \path[fill overzoom image=figures/#1.jpg] (0,0) rectangle (6.5cm,4.5cm);
        \draw[draw=red,thick] (5.5,1.3) -- ((6.5,1.3) -- ((6.5,1.7) -- ((5.5,1.7) -- cycle;
        \draw[draw=blue,thick] (0.1,1.8) -- ((1.1,1.8) -- ((1.1,2.2) -- ((0.1,2.2) -- cycle;
    \end{scope}
}

\newcommand{\arcplot}[1]{
    \begin{scope}
        \clip (0,0) -- (6.5,0) -- (6.5,4.5) -- (0,4.5) -- cycle;
        \path[fill overzoom image=figures/#1.jpg] (0,0) rectangle (6.5cm,4.5cm);
        \draw[draw=red,thick] (2.0,1.8) -- ((3.0,1.8) -- ((3.0,2.2) -- ((2.0,2.2) -- cycle;
        \draw[draw=blue,thick] (0.1,0.8) -- ((1.1,0.8) -- ((1.1,1.2) -- ((0.1,1.2) -- cycle;
    \end{scope}
}

\newcommand{\micro}[1]{
    \begin{scope}
        \clip (0,0) -- (6.5,0) -- (6.5,4.5) -- (0,4.5) -- cycle;
        \path[fill overzoom image=figures/#1.jpg] (0,0) rectangle (6.5cm,4.5cm);
        \draw[draw=blue,thick] (4.8,0.8) -- ((5.8,0.8) -- ((5.8,1.2) -- ((4.8,1.2) -- cycle;
        \draw[draw=red,thick] (0.1,1.8) -- ((1.1,1.8) -- ((1.1,2.2) -- ((0.1,2.2) -- cycle;
    \end{scope}
}

\newcommand{\plotsidebyside}[2]{
    \begin{scope}
        \clip (0,0) -- (3.35,0) -- (3.35,2.25)-- (0.0,2.25) -- cycle;
        \path[fill overzoom image=figures/#2] (0,0) rectangle (3.25,2.25);
    \end{scope}
    \begin{scope}
        \clip (3.35,0) -- (6.5,0) -- (6.5,2.25)-- (3.35,2.25) -- cycle;
        \path[fill overzoom image=figures/#1] (6.5,0) rectangle (3.25,2.25);
    \end{scope}
    \draw[draw=blue,line width=0.3mm] (0,0) -- (3.25,0) -- (3.25,2.25)-- (0.0,2.25) -- cycle;
    \draw[draw=red,line width=0.3mm] (3.35,0) -- (6.5,0) -- (6.5,2.25)-- (3.35,2.25) -- cycle;
}

\small
\setlength\tabcolsep{1pt}
\begin{tabularx}{\linewidth}{cccc}
 Uncorrelated sampling & \citet{abdalla2020screen} & \citet{belcour2021rank1} & \textbf{Ours FSOT}
\\
\begin{tikzpicture}[scale=0.67]
    \chopper{chopper_s1_box_percep/chopper-titan_directlight_random_n1_box_perceptual}
\end{tikzpicture}
&
\begin{tikzpicture}[scale=0.67]
    \chopper{chopper_s1_box_percep/chopper-titan_directlight_Abdalla_n1_box_perceptual}
\end{tikzpicture}
&
\begin{tikzpicture}[scale=0.67]
    \chopper{chopper_s1_box_percep/chopper-titan_directlight_Heitz21_n1_box_perceptual}
\end{tikzpicture}
&
\begin{tikzpicture}[scale=0.67]
    \chopper{chopper_s1_box_percep/chopper-titan_directlight_Ours_n1_box_perceptual}
\end{tikzpicture}
\\
\begin{tikzpicture}[scale=0.67]
    \plotsidebyside{chopper_s1_box_percep/chopper-titan_directlight_random_n1_box_perceptual_b1}{chopper_s1_box_percep/chopper-titan_directlight_random_n1_box_perceptual_b2}
\end{tikzpicture}
&
\begin{tikzpicture}[scale=0.67]
    \plotsidebyside{chopper_s1_box_percep/chopper-titan_directlight_Abdalla_n1_box_perceptual_b1}{chopper_s1_box_percep/chopper-titan_directlight_Abdalla_n1_box_perceptual_b2}
\end{tikzpicture}
&
\begin{tikzpicture}[scale=0.67]
    \plotsidebyside{chopper_s1_box_percep/chopper-titan_directlight_Heitz21_n1_box_perceptual_b1}{chopper_s1_box_percep/chopper-titan_directlight_Heitz21_n1_box_perceptual_b2}
\end{tikzpicture}
&
\begin{tikzpicture}[scale=0.67]
    \plotsidebyside{chopper_s1_box_percep/chopper-titan_directlight_Ours_n1_box_perceptual_b1}{chopper_s1_box_percep/chopper-titan_directlight_Ours_n1_box_perceptual_b2}
\end{tikzpicture}
\\
    \begin{tikzpicture}[scale=0.67]
    \micro{microcity_s1_box_percep/microcity_directlight_random_n1_box_perceptual}
\end{tikzpicture}
&
\begin{tikzpicture}[scale=0.67]
    \micro{microcity_s1_box_percep/microcity_directlight_Abdalla_n1_box_perceptual}
\end{tikzpicture}
&
\begin{tikzpicture}[scale=0.67]
    \micro{microcity_s1_box_percep/microcity_directlight_Heitz21_n1_box_perceptual}
\end{tikzpicture}
&
\begin{tikzpicture}[scale=0.67]
    \micro{microcity_s1_box_percep/microcity_directlight_Ours_n1_box_perceptual}
\end{tikzpicture}
    \\ 
\begin{tikzpicture}[scale=0.67]
    \plotsidebyside{microcity_s1_box_percep/microcity_directlight_random_n1_box_perceptual_b1}{microcity_s1_box_percep/microcity_directlight_random_n1_box_perceptual_b2}
\end{tikzpicture}
&
\begin{tikzpicture}[scale=0.67]
    \plotsidebyside{microcity_s1_box_percep/microcity_directlight_Abdalla_n1_box_perceptual_b1}{microcity_s1_box_percep/microcity_directlight_Abdalla_n1_box_perceptual_b2}
\end{tikzpicture}
&
\begin{tikzpicture}[scale=0.67]
    \plotsidebyside{microcity_s1_box_percep/microcity_directlight_Heitz21_n1_box_perceptual_b1}{microcity_s1_box_percep/microcity_directlight_Heitz21_n1_box_perceptual_b2}
\end{tikzpicture}
&
\begin{tikzpicture}[scale=0.67]
    \plotsidebyside{microcity_s1_box_percep/microcity_directlight_Ours_n1_box_perceptual_b1}{microcity_s1_box_percep/microcity_directlight_Ours_n1_box_perceptual_b2}
\end{tikzpicture}
\end{tabularx}
    \vspace{-3mm}
    \caption{
        Comparison of our perceptual-error optimization against classical uncorrelated pixel sampling and state-of-the-art blue-noise error distribution methods. The top scene is directly lit by an environment map, and the bottom scene has defocus blur that increases the sampling dimensions to four.
    }
    \label{fig:main_rendering_result}
\end{figure*}

For rendering applications, we optimize a point set covering 128$\times$128 pixels that is toroidally tiled over the image. 
In \cref{fig:main_rendering_result}, we compare our point sets against those from prior work on perceptual (\ie blue-noise) error optimization \citep{abdalla2020screen,belcour2021rank1}; we use box reconstruction for a fair comparison. Both scenes are rendered with 1 sample/pixel under direct lighting. The benefit of our approach (rightmost column) is most visible in the top scene, where the specular regions show a much improved error distribution. In the bottom row scene, we use a finite-aperture camera. The zoom-ins show better perceived quality achieved by our method over the state of the art. We provide more comparisons on different scenes in the supplemental material.

\begin{figure}[t!]
    \centering

\definecolor{OutlineColor1}{gray}{0.23}
\definecolor{OutlineColor2}{gray}{0.23}

\newcommand{\OneImage}[1]{
    \begin{scope}
        \clip (0.0,0.0) -- (9.5,0.0) -- (9.5,4.5) -- (0.0,4.5) -- cycle;
        \path[fill overzoom image=figures/#1] (0,0) rectangle (9.5cm,4.5cm);
    \end{scope}
}

\newcommand{\TestVeachImage}[2]{
    \begin{scope}
        \clip (0.0,4.5) -- (9.5,4.5) -- (9.5,7) -- (0.0,7) -- cycle;
        \path[fill overzoom image=figures/veach_mis/#1.jpg] (0,0) rectangle (9.5cm,7cm);
    \end{scope}
    \begin{scope}
        \clip (0.0,0.0) -- (9.5,0.0) -- (9.5,4.5) -- (0.0,4.5) -- cycle;
        \path[fill overzoom image=figures/veach_mis/#2.jpg] (0,0) rectangle (9.5cm,7cm);
    \end{scope}

    \begin{scope}
       \draw[draw=white,ultra thick] (0.0,4.5) -- (9.5,4.5);
       \draw[draw=white,ultra thick] (0.0,2.1) -- (9.5,2.1);
       \draw[draw=white,ultra thick] (2.5,0) -- (2.5,4.5);
      \draw[draw=white,ultra thick] (5,0) -- (5,2.1);
    \end{scope}
}

\newcommand{\VeachCrop}[5]{
    \begin{scope}
        \clip (#1,#2) -- (#3,#2) -- (#3,#4) -- (#1,#4) -- cycle;
        \path[fill overzoom image=figures/veach_mis/#5.jpg] (0,0) rectangle (9.5cm,7cm);
    \end{scope}
}

\newcommand{\VeachImage}[6]{
    \begin{scope}
        \clip (0.0,0.0) -- (2.5,0.0) -- (2.5,7) -- (0.0,7) -- cycle;
        \path[fill overzoom image=figures/veach_mis/#2.jpg] (0,0) rectangle (9.5cm,7cm);
    \end{scope}
    \begin{scope}
        \clip (2.5,2.1) -- (9.5,2.1) -- (9.5,7) -- (2.5,7) -- cycle;
        \path[fill overzoom image=figures/veach_mis/#3.jpg] (0,0) rectangle (9.5cm,7cm);
    \end{scope}
    \begin{scope}
        \clip (2.5,0.0) -- (5,0.0) -- (5,2.1) -- (2.5,2.1) -- cycle;
        \path[fill overzoom image=figures/veach_mis/#5.jpg] (0,0) rectangle (9.5cm,7cm);
    \end{scope}
    \begin{scope}
        \clip (5.0,0.0) -- (9.5,0.0) -- (9.5,2.1) -- (5.0,2.1) -- cycle;
        \path[fill overzoom image=figures/veach_mis/#6.jpg] (0,0) rectangle (9.5cm,7cm);
    \end{scope}

    \begin{scope}
    \end{scope}
}

\small
\hspace*{-2.3mm}
\begin{tabular}{c@{}}
\newcommand{\SplitOne}{2.7}%
\newcommand{\SplitTwo}{4.75}%
\newcommand{\SplitThree}{6.8}%
\begin{tikzpicture}[scale=0.9]
    \VeachCrop{0}{0}{\SplitOne}{7.8}{veach-mis_low_res_directlight_Ours_n4_box_no-perceptual}
    \VeachCrop{\SplitOne}{0}{\SplitTwo}{7.8}{veach-mis_low_res_directlight_Ours_n4_gaussian_no-perceptual_independant_pixels}
    \VeachCrop{\SplitTwo}{0}{\SplitThree}{7.8}{veach-mis_low_res_directlight_Ours_n4_gaussian_no-perceptual}
    \VeachCrop{\SplitThree}{0}{9.5}{7.8}{veach-mis_low_res_directlight_Ours_n4_gaussian_perceptual}
    \begin{scope}[overlay]
        \fill [black,opacity=0.4] (0,7.4) rectangle (9.5,7.8);
        \fill [black,opacity=0.3] (0,7) rectangle (9.5,7.4);
        \draw[draw=black,line width=0.15mm] (\SplitOne,0) -- (\SplitOne,7.8);
        \draw[draw=black,line width=0.15mm] (\SplitTwo,0) -- (\SplitTwo,7.4);
        \draw[draw=black,line width=0.15mm] (\SplitThree,0) -- (\SplitThree,7.4);
        \filldraw[black,ultra thick] (1.35, 7.5) circle (0pt)  node[anchor=base,rotate=0]{\fontsize{8}{6}\selectfont \Outlined[white][OutlineColor1][0.7pt]{Box reconstuction}};
        \filldraw[black,ultra thick] (1.35, 7.12) circle (0pt)  node[anchor=base,rotate=0]{\fontsize{7}{6}\selectfont \Outlined[white][OutlineColor2][0.65pt]{(a)\,Box optimization}};
        %
        \filldraw[black,ultra thick] (3.725, 7.12) circle (0pt)  node[anchor=base,rotate=0]{\fontsize{7}{6}\selectfont \Outlined[white][OutlineColor2][0.65pt]{(b)\,Box\,optim.}};
        \filldraw[black,ultra thick] (5.775, 7.5) circle (0pt)  node[anchor=base,rotate=0]{\fontsize{8}{6}\selectfont \Outlined[white][OutlineColor1][0.7pt]{Gaussian reconstruction}};
        \filldraw[black,ultra thick] (5.775, 7.12) circle (0pt)  node[anchor=base,rotate=0]{\fontsize{7}{6}\selectfont \Outlined[white][OutlineColor2][0.65pt]{(c)\,Gaussian\,opt.}};
        %
        \filldraw[black,ultra thick] (8.15, 7.12) circle (0pt)  node[anchor=base,rotate=0]{\fontsize{7}{6}\selectfont \Outlined[white][OutlineColor2][0.65pt]{(d)\,Gauss.+percep.\,opt.}};
    \end{scope}
\end{tikzpicture}%
\end{tabular}
    \vspace{-2.5mm}
    \caption{
        Optimizing pixel samples for different reconstruction kernels. (a)~When using box reconstruction, the samples for individual pixels can be optimized separately. (b)~Traditionally this optimization is used also when the reconstruction is non-box. (c)~Our framework allows optimizing for the specific kernel used, taking into account overlaps between pixels and showing substantial error reduction. 
        (d)~Additionally taking into account perceptual blur achieves 
        blue-noise error distribution over the image..
    }
    \label{fig:reconstruction_problem_rendering}
\end{figure}
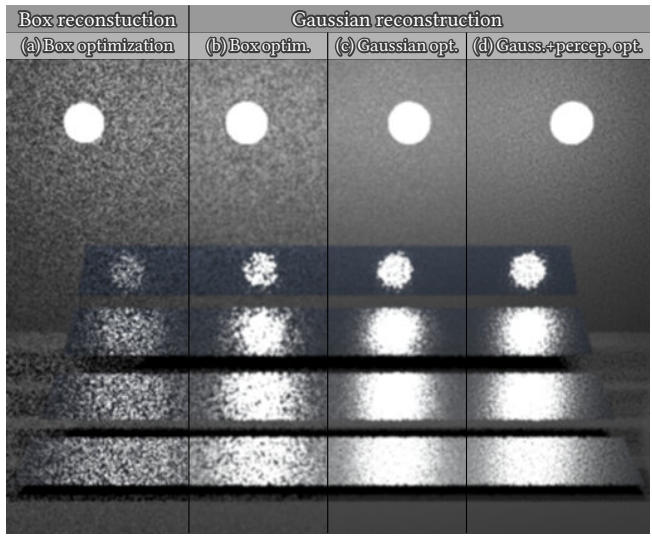

While traditionally point sets are optimized assuming a box pixel-reconstruction kernel, our framework allows optimizing for arbitrary kernels. \Cref{fig:reconstruction_problem_rendering} shows the impact on error distribution while taking into account the reconstruction kernel. On the right, note the substantial improvement in \cref{fig:reconstruction_problem_rendering}c over \cref{fig:reconstruction_problem_rendering}b, due to specially optimizing for the Gaussian reconstruction kernel used. Additionally accounting for perceptual blur further pushes the error distribution toward high frequencies (\cref{fig:reconstruction_problem_rendering}d).
Another comparison against the method of \citet{belcour2021rank1} using box and Gaussian pixel reconstruction can be found in the supplemental document. It shows the importance of optimization not only for perceptual blur but also for the pixel-reconstruction kernel.

\subsection{Algorithmic complexity and performance}

The bottleneck of our optimization is the sorting of $m$ projected optimization points and $c\cdot m$ density-sample points (where $c = \text{const}$), with complexity $O(m\,\mathrm{log}(m))$ per iteration. The number of classes and subclasses has no direct impact on complexity, although in practice increasing the number of optimization objectives can impact the convergence speed of gradient descent. The memory consumption of our algorithm is linear in the total number of \mbox{optimization points $\PointCount$.}

For 4096 points, single-class GPU optimization takes 40\,sec, 3-class takes 59\,sec, and 7-class takes 71\,sec. For 262,144 points, single-class takes 3840\,sec (2000 iterations), 3-class takes 4325\,sec (2500 iterations), and 7-class takes 4370\,sec (3000 iterations). The added cost of increasing the number of classes is moderate. The reason is that, while more classes require more optimization iterations to obtain high quality, the time per iteration is lower as fewer points are optimized at once (since one subclass it optimized per iteration). With this in mind, it is possible to imagine a more efficient optimization, \eg utilizing a data structure to pre-order the points before projection and then using a sorting algorithm that takes advantage of this pre-ordering. One can also imagine relaxing the constraints on the Wasserstein equations to perform local rather than global optimizations. By computing several Wasserstein distances on subsets of the domain, it is possible to approximate the full distance with fewer points in each ``sub-distance''. Because of the complexity of these operations, reducing the number of points would speed up the computation at the cost of a looser error bound.

\section{Conclusion}

We develop a theoretical point optimization framework designed for handling large numbers of objectives. Specifying these objectives for a given application can be tedious if done manually. Prior methods~\cite{wei2010multiclass,qin2017wasserstein} have overlooked this issue as they target applications with fewer objectives.

We devise a principled framework for point optimization that can handle large numbers of objectives. We introduce the notion of a subclass which adds a level of granularity by specifying an objective for a subset of points in a class. Our framework scales to such a large number of objectives because, theoretically, the achievable quality does not depend on the number of objectives but on the amount of overlap between them and the difference in target distributions. The memory footprint of our stochastic gradient-descent optimization is negligible as we optimize a single subclass per iteration. We demonstrate a variety of applications, also formalizing perceptual-error optimization as a multi-class problem.

\Paragraph{Limitations}

Our multi-class Wasserstein barycenter objective has a fully integral form, which allows us to leverage stochastic optimization and achieve scalability. However, optimizing for a single objective per iteration can yield noisy gradients and slow down convergence toward the sought barycenter. Our point-sampling of non-uniform distributions exacerbates the issue by adding more noise to the gradients.

Wherever functions of different classes overlap, points are implicitly optimized toward a barycenter of the corresponding target distributions. Some applications require a union of point subsets to follow a \emph{mixture} of the targets instead. A notable example is color stippling where the base targets are the distributions of the individual color channels. Our framework requires specifying mixture targets explicitly via dedicated classes.

\Paragraph{Future work}

Our optimization can benefit from analytic target-distribution projection and informed choices of projection axes that allows tailoring application-specific samplers. 
A more advanced optimizer could achieve better local minima than stochastic gradient descent. While enabling efficient optimization, the sliced Wasserstein barycenter we use may not yield a good distribution interpolation~\citep{Bonneel:Sliced,Bonneel2013SlicedWB}. Efficient optimization of the regular Wasserstein barycenter is an interesting direction for future investigation.

The Wasserstein distance provides a convenient integration-error bound as it is amenable to gradient-based minimization. However, the tightness of that bound is not well understood, especially in relation to the discrepancy-based bound given by the Koksma-Hlawka inequality. Exploring this relation could help better understand the optimization manifolds for future sampling patterns. Another interesting investigation would be the efficient minimization of discrepancy metrics.

\begin{acks} 
We thank all the anonymous reviewers for their helpful comments in shaping the final version of the paper.
We thank the following for scenes used in our experiments: julioras3d (\textsc{chopper-titan}), Mikael Hvidtfeldt Christensen (\textsc{structuresynth}), Greyscalegorilla (\textsc{vw-van}) and Eric Veach (\textsc{Veach-mis}). We also thanks Sponchia for the elephants image.
\end{acks}

\bibliographystyle{ACM-Reference-Format}
\bibliography{paper}


\appendix

\section{1D Wasserstein distance derivative}
\label{sec:WasersteinDerivative}

Here we derive the derivative of the 1D Wasserstein distance which has an analytic solution~\citep{Rachev:MassTransportation}:
\begin{equation}
    \label{eq:WassersteinDistance1D}
    W_p^p(\OptMeasure,\TargetMeasure) \, = \int_0^\infty \! \big| \, F^{-1}_\OptMeasure(\Point) - F^{-1}_\TargetMeasure(\Point) \big|^{\,p} \dif\Point,
\end{equation}
where $F^{-1}_\OptMeasure~$ and $F^{-1}_\TargetMeasure$ are the measures' inverse cumulative distribution functions (CDFs). We are specifically interested in the case where $p=2$ and one of the measures represents a 1D point set $\PointSet = \{ \Point_i \}_{i=1}^\PointCount$. For this case we have
\begin{align}
    W^2_2(\PointSet,\TargetMeasure)
        &= \int_0^1 \big[F^{-1}_\PointSet(\Point) - F^{-1}_\TargetMeasure(\Point)\big]^2 \dif\Point 
        = \sum_{i=1}^\PointCount \int_{\frac{i-1}{\PointCount}}^{\frac{i}{\PointCount}} \big(\Point_i - F^{-1}_\TargetMeasure(\Point)\big)^2 \dif\Point.\nonumber
\end{align}
We want to differentiate this distance \wrt every point $\Point_i$. Only one of the integrals depends on each $\Point_i$, making the differentiation of its convex integrand easy:
\begin{align}
    \frac{\mathrm{d}}{\mathrm{d} \Point_i} W^2_2(\PointSet,\TargetMeasure)
        &= \!\int_{\frac{i-1}{\PointCount}}^{\frac{i}{\PointCount}} \frac{\mathrm{d}}{\mathrm{d} \Point_i} \big(\Point_i^2 - 2 \Point_i F^{-1}_\TargetMeasure(\Point) + (F^{-1}_\TargetMeasure(\Point))^2\big) \,\dif\Point \\
        \nonumber
        &= \!\int_{\frac{i-1}{\PointCount}}^{\frac{i}{\PointCount}} \!\! 2 (\Point_i - F^{-1}_\TargetMeasure(\Point)) \,\dif\Point
        = \, 2 \frac{\Point_i}{\PointCount} - 2 \! \int_{\frac{i-1}{\PointCount}}^{\frac{i}{\PointCount}} \!\! F^{-1}_\TargetMeasure(\Point) \dif\Point.
\end{align}
The integral of the inverse target CDF is simply the average point inside the $i$\textsuperscript{th} region of the target density with mass $1/\PointCount$, multiplied by $1/\PointCount$. The resulting derivative is thus similar to offset used by \citet{paulin2020sliced}; the difference is the above scaling factor of $2/\PointCount$ and that they take the median target point (instead of the mean).

\section{Gradient estimation and point offsets}
\label{sec:GradientEstimationPointOffsets}

The 1D optimization step involves offsetting each projected point $\Point_i^\theta$ along the (negative) derivative of the Wasserstein distance \wrt the point's position:
\begin{align}
    \label{eq:GradientOptimization}
    \Point_i^\theta
        \,=\, \Point_i^\theta - \eta \cdot \gamma_i^\theta \cdot \underbracket[0.5pt][1.4pt]{\frac{\mathrm{d}}{\mathrm{d} \Point_i} W^p_p(\PointSetExt_{\ClassFunc > \Slice}^\theta,\TargetMeasure_{\ClassFunc > \Slice}^\theta)}_{\Delta_i^\theta},
\end{align}
where $\eta$ is a step-size parameter (a.k.a.\ learning rate) and $\gamma$ the offset scaling factor (\cref{sec:Stochastic_gradient-descent_optimization}). In \cref{sec:WasersteinDerivative} above we derive the derivative for the semi-discrete 2-Wasserstein distance for the case where both distributions are normalized. In our case they have reduced mass due to filtering, which can be compensated by simply scaling the derivative by the relative number of selected optimization points $m/\PointCount = \big| \PointSetExt_{\ClassFunc > \Slice} \big|/\PointCount$:
\begin{align}
    \label{eq:Derivative}
    \Delta_i^\theta \,=\, \frac{m}{\PointCount} \left[ 2 \frac{\Point_i^\theta}{m} - 2 \! \int_{\frac{i-1}{m}}^{\frac{i}{m}} \!\! F^{-1}_{\ProjectedTargetMeasure}(\Point) \dif\Point \right],
\end{align}
where the semi-discrete derivative in the parentheses is computed \wrt normalized distributions.

\Paragraph{Numerical gradient estimation}

The partial derivative step~\eqref{eq:GradientOptimization} requires computing the inverse CDF of the projected target distribution $\ProjectedTargetMeasure$. 
In practice, we use $c\times\PointCount$ points to better approximate the target distribution.
First, all points are uniformly binned in \PointCount bins.
The inverse CDF then adaptively changes the bin length according to the target distribution.

The integral term in~\cref{eq:Derivative} corresponds to the average location of the points within a certain interval of the inverse CDF:
%
$
    b_i^\theta = \nicefrac{1}{c}\sum_{j=(i-1) c}^{i c} y_j^\theta,
$
%
where the projected target samples $y_j^\theta$ are sorted.
This gives the \emph{average} location per $i$-th bin.
Computing the offset $\Delta_i^\theta$ then involves sorting $\Point_i^\theta$ and pairwise matching them with the bin values $b_i^\theta$.

\Paragraph{Gradient scaling factor}
The scaling factor $\gamma_i$~\eqref{eq:GradientOptimization} is simply the relative change in the length of the $i$-th bin:
\begin{align}
    \label{eq:offset_correction_factor}
    \gamma_i^\theta \,=\,
    \frac{\text{Average bin length}}{\text{Length of bin $i$}}
    =
    \frac{\nicefrac{(F^{-1}_{\ProjectedTargetMeasure}(1)-F^{-1}_{\ProjectedTargetMeasure}(0))}{m}}{F^{-1}_{\ProjectedTargetMeasure}(\nicefrac{i}{m})-F^{-1}_{\ProjectedTargetMeasure}(\nicefrac{(i-1)}{m})}
\end{align}
This scaling factor helps avoid the alignments shown in~\cref{fig:pointsets_comparison}c by scaling the gradients (offsets) for the projected non-uniform target density.

\section{Wasserstein integration-error bound}
\label{sec:ErrorBoundDerivation}

Here we provide a derivation of the integration error bound shown by \citet{paulin2020sliced}. Consider a continuous function $f : \Hypercube \! \rightarrow \! \RealsPlus$ on the hypercube $\Hypercube$ with Lipschitz constant $L_f$ \mbox{such that, $\forall x,y \in \Hypercube$},
\begin{equation}
    \label{eq:LipschitzInequality}
    |f(x) - f(y)| \leq L_f \|x - y\|.
\end{equation}
Let $\gamma \in \Gamma(\OptMeasure,\TargetMeasure)$ be a joint measure whose marginals $\OptMeasure$ and $\TargetMeasure$ are measures on the unit hypercube $\Hypercube$. Integrating both sides of \cref{eq:LipschitzInequality} w.r.t.\ $\gamma$, and then using $\big|\!\int \! g \big| \leq \int \! |g|$, yields
\begin{align}
    \int_{\Hypercube^2} |f(x) - f(y)| \, \dif \gamma(x,y) &\leq L_f \! \int_{\Hypercube^2} \|x - y\| \, \dif \gamma(x,y) \\
    \label{eq:ErrorBoundDeriv1}
    \left| \int_{\Hypercube^2} \big[ f(x) - f(y) \big] \, \dif \gamma(x,y) \right| &\leq L_f \! \int_{\Hypercube^2} \|x - y\| \, \dif \gamma(x,y).
\end{align}
We expand the left side of \cref{eq:ErrorBoundDeriv1} into two integrals and simplify each by marginalizing the product measure; the bound on the right is tightened by taking the infimum over all valid joint measures $\gamma$:
\begin{equation}
    \label{eq:ErrorBoundAppendix}
    \ScaleMath{0.93}{
        \!\left| \int_\Hypercube \!\!\! f(x) \,\dif\OptMeasure(x) - \! \int_\Hypercube \!\!\! f(x) \,\dif\TargetMeasure(x) \right| \leq L_f \! \underbracket[0.5pt][1.2pt]{ \inf_{\gamma \in \Gamma(\OptMeasure,\TargetMeasure)} \int_{\Hypercube^2} \!\! \|x - y\| \,\dif \gamma(x,\!y) }_{W(\OptMeasure,\TargetMeasure)},
    }\\[-1mm]
\end{equation}
where $W(\OptMeasure,\TargetMeasure)$ is the Wasserstein distance between $\OptMeasure$ and $\TargetMeasure$. The resulting inequality provides a numerical integration bound when $\OptMeasure$ is a Dirac point-mass measure, \ie a point set. 

\section{Reconstruction-error bound}
\label{sec:WeightedErrorBoundDerivation}

We build on \cref{sec:ErrorBoundDerivation} to derive an error bound for integrands of the form $\ClassFunc(x) f(x)$, where $\ClassFunc$ is an analytically known function. As in \cref{sec:ErrorBoundDerivation}, our derivations use general probability measures $\OptMeasure$ and $\TargetMeasure$, but for our application we are specifically interested in the case where $\OptMeasure$ is a Dirac point-mass measure, \ie a point set. 

We begin by expressing $\ClassFunc(x)$ and $\ClassFunc(y)$ in the error as integrals over corresponding indicator functions, then swap the integration order using Fubini's theorem:
\begin{align}
    \label{eq:SlicedWeightDerivation0}
    &\!\!\left |\int_\Hypercube \!\! \ClassFunc(x) f(x) \,\dif\OptMeasure(x) - \int_\Hypercube \!\! \ClassFunc(y) f(y) \,\dif\TargetMeasure(y)\right |\\
    \nonumber
    &\;=\ScaleMath{0.93}{
        \bigg | \! \int_\Hypercube \! \underbracket[0.5pt][1.2pt]{\int_{\Reals} \!\! \Indicator_{[0,\!\ClassFunc(x)\!]\!}(z) \dif z}_{\ClassFunc(x)}  f(x) \, \dif\OptMeasure(x) - \! \int_\Hypercube \! \underbracket[0.5pt][1.2pt]{\int_{\Reals} \!\! \Indicator_{[0,\!\ClassFunc(y)\!]\!}(z) \dif z}_{\ClassFunc(y)} f(y) \, \dif\TargetMeasure(y) \bigg |
    }\\
    \label{eq:SlicedWeightDerivation1}
    &\;=\ScaleMath{0.91}{
        \left | \int_{\Reals} \! \left[ \int_\Hypercube \!\! \Indicator_{[0,\!\ClassFunc(x)\!]\!}(z) f(x) \, \dif\OptMeasure(x) - \! \int_\Hypercube \!\! \Indicator_{[0,\!\ClassFunc(y)\!]\!}(z) f(y) \,\dif\TargetMeasure(y) \right] \! \dif z\right |.
    }\!
\end{align}
Next, note that due the following identity for any $x \in \Hypercube$ and $z \in \Reals$:
\begin{equation}
    \ScaleMath{0.93}{
        \Indicator_{[\!0,\ClassFunc(x)\!]}(z) = \Indicator_{[z,\infty]}(\ClassFunc(x)) = \Indicator_{\{x' \in \Hypercube \mid \ClassFunc(x') > z\}}(x) =\vcentcolon \Indicator_{\Hypercube_{\ClassFunc>z}\!}(x),
    }
\end{equation}
the indicator function effectively restricts the integration to the region $\Hypercube_{\ClassFunc > z}$ where $\ClassFunc(\cdot) > z$. Plugging this identity into \cref{eq:SlicedWeightDerivation1} and then using $\big|\!\int \! g \big| \leq \int \! |g|$, we get
\begin{align}
    &=\left | \int_{\Reals} \! \left[ \int_\Hypercube \!\!\! \Indicator_{\Hypercube_{\ClassFunc > z}\!}(x) f(x) \,\dif\OptMeasure(x) - \! \int_\Hypercube \!\!\! \Indicator_{\Hypercube_{\ClassFunc > z}\!}(y) f(y) \,\dif\TargetMeasure(y) \right] \! \dif z \right| \\
    &\leq \int_{\Reals} \left | \, \int_\Hypercube \!\!\! \Indicator_{\Hypercube_{\ClassFunc > z}\!}(x) f(x) \,\dif\OptMeasure(x) - \! \int_\Hypercube \!\!\! \Indicator_{\Hypercube_{\ClassFunc > z}\!}(y) f(y) \,\dif\TargetMeasure(y) \, \right| \dif z \\
    &= \int_{\Reals} \, \left | \int_\Hypercube \!\! f(x) \,\dif\OptMeasure_{\ClassFunc > z}(x) - \int_\Hypercube \!\! f(y) \,\dif\TargetMeasure_{\ClassFunc > z}(y) \right | \dif z,
\end{align}
where $\OptMeasure_{\ClassFunc > z}$ and $\TargetMeasure_{\ClassFunc > z}$ are the measures $\OptMeasure$ and $\TargetMeasure$ restricted to the region $\Hypercube_{\ClassFunc > z}$. We can now apply \cref{eq:ErrorBoundAppendix} to the absolute error in the outer integral, obtaining a bound for the expression in \cref{eq:SlicedWeightDerivation0}:
\begin{align}
    \label{eq:WeightSlicedBound}
    \!\ScaleMath{0.95}{
        \!\! \left |\int_\Hypercube \!\!\!\! \ClassFunc(x) f(x) \dif\OptMeasure(x) - \! \int_\Hypercube \!\!\!\! \ClassFunc(y) f(y) \dif\TargetMeasure(y)\right |
        \leq L_f \!\! \int_{\Reals} \!\! W(\OptMeasure_{\ClassFunc>z},\!\TargetMeasure_{\ClassFunc>z}) \dif z.
    }\!\!
\end{align}
It is important to note that for the Wasserstein distance to work, the measures $\OptMeasure_{\ClassFunc>z}$ and $\TargetMeasure_{\ClassFunc>z}$ must have equal masses in the hypercube subset corresponding to each valid slicing $\ClassFunc>z$. In other words, we need $\OptMeasure(\Hypercube_{\ClassFunc>z}) = \TargetMeasure(\Hypercube_{\ClassFunc>z})$, or equivalently, $\OptMeasure_{\ClassFunc>z}(\Hypercube) = \TargetMeasure_{\ClassFunc>z}(\Hypercube)$, for all $z \in [0, \max w(.)]$.

\end{document}



\title{Scalable multi-class sampling via filtered sliced optimal transport: Supplemental document}

\author{Corentin Sala\"un}
\affiliation{%
  \institution{Max-Planck-Institut f{\"u}r Informatik}
  \city{Saarbr{\"u}cken}
  \country{Germany}
}

\author{Iliyan Georgiev}
\affiliation{%
  \institution{Autodesk}
  \country{United Kingdom}
}

\author{Hans-Peter Seidel}
\affiliation{%
  \institution{Max-Planck-Institut f{\"u}r Informatik}
  \city{Saarbr{\"u}cken}
  \country{Germany}
}

\author{Gurprit Singh}
\affiliation{%
  \institution{Max-Planck-Institut f{\"u}r Informatik}
  \city{Saarbr{\"u}cken}
  \country{Germany}
}

\begin{abstract}

In this document we propose the derivation of the 1D 1-Wasserstein distance and additional results to the main document. You can find a second image of stippling in color for a lighthouse image. A monochrome stippling of elephants and an animation stippling example. We finally present two additional comparison for the rendering.

\end{abstract}


\maketitle

\section{1D 1-Wasserstein distance derivative}
\label{sec:1WasersteinDerivative}

Here we derive the derivative of the 1D Wasserstein distance for which has a closed-form exists~\citep{Rachev:MassTransportation}:
\begin{equation}
    \label{eq:WassersteinDistance1D}
    W_p^p(\OptMeasure,\TargetMeasure) \, = \int_0^\infty \! \big| \, F^{-1}_\OptMeasure(\Point) - F^{-1}_\TargetMeasure(\Point) \big|^{\,p} \dif\Point,
\end{equation}
where $F^{-1}_\OptMeasure~$ and $F^{-1}_\TargetMeasure$ are the measures' inverse CDFs. We are specifically interested in the case where $p=1$ and one of the measures represents a 1D point set $\PointSet = \{ \Point_i \}_{i=1}^\PointCount$. For this case we have
\begin{align}
    W^1_1(\PointSet,\TargetMeasure)
        &= \int_0^1 \big|(F^{-1}_\PointSet(\Point) - F^{-1}_\TargetMeasure(\Point)\big| \dif\Point \\
        &= \sum_{i=1}^\PointCount \int_{\frac{i-1}{\PointCount}}^{\frac{i}{\PointCount}} \big|\Point_i - F^{-1}_\TargetMeasure(\Point)\big| \dif\Point.
\end{align}
We want to differentiate this distance \wrt every point $\Point_i$. Only one of the integrals depends on each $\Point_i$:
\begin{align}
    \frac{\mathrm{d}}{\mathrm{d} \Point_i} W^1_1(\PointSet,\TargetMeasure)
        &= 
        \begin{dcases}
            \!\int_{\frac{i-1}{\PointCount}}^{\frac{i}{\PointCount}} 1 \,\dif\Point ,& \text{if } \big(\Point_i - F^{-1}_\TargetMeasure(\Point)\big)\geq 0\\
            \!\int_{\frac{i-1}{\PointCount}}^{\frac{i}{\PointCount}} -1 \,\dif\Point ,              & \text{otherwise}
        \end{dcases}
\end{align}
The integral of the inverse target CDF is simply the average point inside the $i$\textsuperscript{th} region of the target density with mass $1/\PointCount$, multiplied by $1/\PointCount$. This derivative is only dependant of the sign of the difference $\big(\Point_i - F^{-1}_\TargetMeasure(\Point)\big)$ making it less attractive in a gradient descent algorithm.

\section{Additional results results}

\begin{figure}[t!]
    \centering

\newcommand{\PlotSplit}[4]{
    \begin{scope}
        \clip (0,0) -- (1.625,0) -- (1.625,4.5) -- (0.0,4.5) -- cycle;
        \path[fill overzoom image=figures/#1] (0,0) rectangle (6.5cm,4.5cm);
    \end{scope}
    \begin{scope}
        \clip (1.625,0) -- (3.25,0) -- (3.25,4.5) -- (1.625,4.5) -- cycle;
        \path[fill overzoom image=figures/#2] (0,0) rectangle (6.5cm,4.5cm);
    \end{scope}
    \begin{scope}
        \clip (3.25,0) -- (4.875,0) -- (4.875,4.5) -- (3.25,4.5) -- cycle;
        \path[fill overzoom image=figures/#3] (0,0) rectangle (6.5cm,4.5cm);
    \end{scope}
    \begin{scope}
        \clip (4.875,0) -- (6.5,0) -- (6.5,4.5) -- (4.875,4.5) -- cycle;
        \path[fill overzoom image=figures/#4] (0,0) rectangle (6.5cm,4.5cm);
    \end{scope}
    \draw[draw=black,thick] (1.625,0) -- (1.625,4.5);
    \draw[draw=black,thick] (3.25,4.5) -- (3.25,0.0);
    \draw[draw=black,thick] (4.875,0) -- (4.875,4.5);
}

\small
\hspace*{-2mm}
\setlength\tabcolsep{1pt}

\begin{tikzpicture}[scale=1.3]
    \PlotSplit{veach-mis_box_gaussian/veach-mis_low_res_directlight_Ours_n1_box_perceptual}{veach-mis_box_gaussian/veach-mis_low_res_directlight_Heitz21_n1_box_perceptual}{veach-mis_box_gaussian/veach-mis_low_res_directlight_Ours_n1_gaussian_perceptual}{veach-mis_box_gaussian/veach-mis_low_res_directlight_Heitz21_n1_gaussian_perceptual}
    \begin{scope}
        \filldraw[black,ultra thick] (1.8,4.6) circle (0pt) node[anchor=base,rotate=0]{ \,Box reconstruction};
        %
        \filldraw[black,ultra thick] (5.0, 4.6) circle (0pt) node[anchor=base,rotate=0]{Gaussian reconstruction};
        %
        \filldraw[black,ultra thick] (0.8,-0.25) circle (0pt) node[anchor=base,rotate=0]{\textbf{Ours}};
        \filldraw[black,ultra thick] (2.4,-0.25) circle (0pt) node[anchor=base,rotate=0]{\citeauthor{belcour2021rank1}};
        \filldraw[black,ultra thick] (4.0,-0.25) circle (0pt) node[anchor=base,rotate=0]{\textbf{Ours}};
        \filldraw[black,ultra thick] (5.65,-0.25) circle (0pt) node[anchor=base,rotate=0]{\citeauthor{belcour2021rank1}};
    \end{scope}
\end{tikzpicture}
     \vspace{-6mm}
     \caption{
         Comparison between our perceptual error optimization and the blue-noise error distribution method of \citet{belcour2021rank1} using box and Gaussian pixel-reconstruction kernels.
         Our method achieves better quality in both cases.
     }
     \label{fig:box-gaussian_rendering}
 \end{figure}

\Paragraph{Rendering comparisons}

\newtext{
We show two additional results for rendering. \Cref{fig:box-gaussian_rendering} shows the comparison of our method with \citet{belcour2021rank1} for rendering at 1 sample per pixel. On the right the comparison shows box reconstruction filter and on the right gaussian filter. We can see that our method performs better in both cases but this difference is more important for the gaussian filter. Contrary to \citet{belcour2021rank1} our method can be optimized for both the perceptual and the reconstruction filter. This difference leads to the improvement visible here.
}

\newtext{
\Cref{fig:tilled_PS_buddha} shows two additional scenes for the integration error distribution comparing an uncorrelated rendering, \citet{abdalla2020screen} , \citet{belcour2021rank1} and our method. For this method we have displayed an image showing a power spectrum per tile in order to show the spectral distribution of the noise. We can see that our method results in the best noise distribution as it gets the fewest low frequencies.
}

\Paragraph{Animated stippling}
\Cref{fig:rotating_logo} shows another example of stippling but across frames. 
Here we encode an entire sequence of 10 frames in a single point set. 
Each frame corresponds to one class with a trapezoidal function and a target distribution set to greyscale frame. 
\newtext{There 75\% of the points overlap between two consecutive frames leading to a large overlap in the class functions. }
That is, each frame is composed of 1,000 points and 250 of them are changed in the next frame. We have displayed 5 frames of the animation in \cref{fig:rotating_logo} as well as the visualization of total point-set.
This example shows the ability of the method to adapt the distribution of the points to jointly satisfy \newtext{multiple objectives with overlap with different target distributions.
A full animations is also available to better appreciate the multi-class nature of the problem.}

\Paragraph{Stippling}

\Cref{fig:stippling} shows another example of color stippling for a lighthouse image with 40 000 points. The first line shows the complete result and the second one shows some classes.

\Cref{fig:stippling_elephants} is an example of monochrome stippling for an image of elephants. 100 000 points are used to generate this image.

\begin{figure*}[t!]

\newcommand{\plotJPG}[1]{
    \begin{tikzpicture}[scale=0.63]
        \begin{scope}
            \clip (0,0) -- (4.5,0) -- (4.5,4.5) -- (0,4.5) -- cycle;
            \path[fill overzoom image=figures/#1.jpg] (0,0) rectangle (5.5cm,3.5cm);
        \end{scope}
    \end{tikzpicture}
}

\newcommand{\plotPNG}[1]{
    \begin{tikzpicture}[scale=0.63]
        \begin{scope}
            \clip (0,0) -- (4.5,0) -- (4.5,4.5) -- (0,4.5) -- cycle;
            \path[fill overzoom image=figures/#1.png] (0,0) rectangle (4.5cm,4.5cm);
            \draw (0,0) -- (4.5,0) -- (4.5,4.5) -- (0,4.5) -- cycle;
        \end{scope}
    \end{tikzpicture}
}

\small
\hspace*{-1.5mm}
\setlength\tabcolsep{1pt}
\begin{tabularx}{\linewidth}{ccccccc}
\plotJPG{annimation_stippling/frame_1}
&
\plotJPG{annimation_stippling/frame_5}
&
\plotJPG{annimation_stippling/frame_9}
&
\plotJPG{annimation_stippling/frame_14}
&
\plotJPG{annimation_stippling/frame_18}
&
&
\plotJPG{annimation_stippling/full}
\\
Frame 1 & Frame 5 & Frame 9 & Frame 14 & Frame 18 & \hspace{1.5mm} & Full point set
\end{tabularx}
    \vspace{-2.5mm}
    \caption{
        Stippling of an 18-frame animation using 5,500 points. Each frame shows the 1,000 points inside an index window that moves by 250 points between consecutive frames; every point thus participates in four frames. Please see the supplemental animation to better appreciate the stippling motion.
    }
    \label{fig:rotating_logo}
\end{figure*}


 \begin{figure*}[t!]
    \centering

\newcommand{\OneImage}[1]{
    \begin{scope}
        \clip (0,0)-- (0.0,4.65) -- (8.75,4.65) -- (8.75,0.0) -- cycle;
        \path[fill overzoom image=figures/#1] (0,0) rectangle (8.5,4.5);
    \end{scope}
    \draw[ thick] (0,0)-- (0.0,4.5) -- (8.5,4.5) -- (8.5,0.0) -- cycle;
}

\newcommand{\SlantedSplitImages}[4]{
    \begin{scope}
        \clip (0,0)-- (0.0,4.5) -- (2.25,4.5) -- (2.0,0.0) -- cycle;
        \path[fill overzoom image=figures/#1] (0,0) rectangle (8.5,4.5);
    \end{scope}
    \begin{scope}
        \clip (2.0,0.0) -- (2.25,4.5) -- (4.375,4.5) -- (4.125,0.0) -- cycle;
        \path[fill overzoom image=figures/#2] (0,0) rectangle (8.5,4.5);
    \end{scope}
    \begin{scope}
        \clip (4.125,0.0)-- (4.375,4.5) -- (6.5,4.5) -- (6.25,0.0) -- cycle;
        \path[fill overzoom image=figures/#3] (0,0) rectangle (8.5,4.5);
    \end{scope}
    \begin{scope}
        \clip (6.25,0.0)-- (6.5,4.5) -- (8.5,4.5) -- (8.5,0.0) -- cycle;
        \path[fill overzoom image=figures/#4] (0,0) rectangle (8.5,4.5);
    \end{scope}
    \draw[draw=black,thick] (2.25,4.5) -- (2.0,0.0);
    \draw[draw=black,thick] (4.375,4.5) -- (4.125,0.0);
    \draw[draw=black,thick] (6.5,4.5) -- (6.25,0.0);
    \draw[draw=white,thin] (0,0.5) -- (1,0.5);
    \draw[ thick] (0,0)-- (0.0,4.5) -- (8.5,4.5) -- (8.5,0.0) -- cycle;
}

\small
\hspace*{-2mm}
\begin{tabular}{c@{}}
    \begin{tikzpicture}[scale=1.6]
        \OneImage{stipling/groix/groix_chat_cmyb.png}
    \end{tikzpicture}
\\
    \hspace*{-0.2cm}\begin{tikzpicture}[scale=1.6]
        \SlantedSplitImages{stipling/groix/groix_chat_y.png}{stipling/groix/groix_chat_my.png}{stipling/groix/groix_chat_cm.png}{stipling/groix/groix_chat_c.png}
        \begin{scope}
            \filldraw[white,ultra thick] (0.3, 4.4) circle (0pt) node[anchor=north,rotate=0] 
            {MC \tiny};
            \filldraw[white,ultra thick] (8.0, 4.4) circle (0pt) node[anchor=north,rotate=0] 
            {Ours \tiny};
        \end{scope}
    \end{tikzpicture}
\end{tabular} 
     \vspace{-2.5mm}
     \caption{
         Stippling image of a lighthouse using 4 colors (CMYK). The complete point-set is composed of 40 000 points and is visible on the top. On the bottom, we display some optimized classes, either individual colors or combinations. The complete problem is composed of 14 classes.
     }
     \label{fig:stippling}
 \end{figure*}
 \begin{figure*}[t!]
    \centering

\newcommand{\HighDimExempleTemplate}[1]{
    \begin{scope}
        \clip (0,0) -- (6.5,0) -- (6.5,4.5) -- (0,4.5) -- cycle;
        \path[fill overzoom image=figures/#1.jpg] (0,0) rectangle (6.5cm,4.5cm);
        
    \end{scope}
}

\newcommand{\chopper}[1]{
    \begin{scope}
        \clip (0,0) -- (6.5,0) -- (6.5,10.5) -- (0,10.5) -- cycle;
        \path[fill overzoom image=figures/#1.jpg] (0,0) rectangle (6.5cm,10.5cm);
    \end{scope}
}

\newcommand{\arcplot}[1]{
    \begin{scope}
        \clip (0,0) -- (6.5,0) -- (6.5,4.5) -- (0,4.5) -- cycle;
        \path[fill overzoom image=figures/#1.jpg] (0,0) rectangle (6.5cm,4.5cm);
        \draw[draw=red,thick] (2.0,1.8) -- ((3.0,1.8) -- ((3.0,2.2) -- ((2.0,2.2) -- cycle;
        \draw[draw=blue,thick] (0.1,0.8) -- ((1.1,0.8) -- ((1.1,1.2) -- ((0.1,1.2) -- cycle;
    \end{scope}
}

\newcommand{\micro}[1]{
    \begin{scope}
        \clip (0,0) -- (6.5,0) -- (6.5,4.5) -- (0,4.5) -- cycle;
        \path[fill overzoom image=figures/#1.jpg] (0,0) rectangle (6.5cm,4.5cm);
    \end{scope}
}

\newcommand{\plotsidebyside}[2]{
    \begin{scope}
        \clip (0,0) -- (3.35,0) -- (3.35,2.25)-- (0.0,2.25) -- cycle;
        \path[fill overzoom image=figures/#2] (0,0) rectangle (3.25,2.25);
    \end{scope}
    \begin{scope}
        \clip (3.35,0) -- (6.5,0) -- (6.5,2.25)-- (3.35,2.25) -- cycle;
        \path[fill overzoom image=figures/#1] (6.5,0) rectangle (3.25,2.25);
    \end{scope}
    \draw[draw=blue,line width=0.3mm] (0,0) -- (3.25,0) -- (3.25,2.25)-- (0.0,2.25) -- cycle;
    \draw[draw=red,line width=0.3mm] (3.35,0) -- (6.5,0) -- (6.5,2.25)-- (3.35,2.25) -- cycle;
}

\small
\setlength\tabcolsep{1pt}
\begin{tabularx}{\linewidth}{cccc}
 Uncorrelated sampling & \citet{abdalla2020screen} & \citet{belcour2021rank1} & \textbf{Ours FSOT}
\\
\begin{tikzpicture}[scale=0.67]
    \chopper{tiled_PS/buddha_directlight_directlighting_n1_box_perceptual}
\end{tikzpicture}
&
\begin{tikzpicture}[scale=0.67]
    \chopper{tiled_PS/buddha_directlight_Abdalla_n1_box_perceptual}
\end{tikzpicture}
&
\begin{tikzpicture}[scale=0.67]
    \chopper{tiled_PS/buddha_directlight_Heitz21_n1_box_perceptual}
\end{tikzpicture}
&
\begin{tikzpicture}[scale=0.67]
    \chopper{tiled_PS/buddha_directlight_Ours_n1_box_perceptual}
\end{tikzpicture}
\\
\begin{tikzpicture}[scale=0.67]
    \chopper{tiled_PS/buddha_directlight_directlighting_n1_box_perceptual_spectrum}
\end{tikzpicture}
&
\begin{tikzpicture}[scale=0.67]
    \chopper{tiled_PS/buddha_directlight_Abdalla_n1_box_perceptual_spectrum}
\end{tikzpicture}
&
\begin{tikzpicture}[scale=0.67]
    \chopper{tiled_PS/buddha_directlight_Heitz21_n1_box_perceptual_spectrum}
\end{tikzpicture}
&
\begin{tikzpicture}[scale=0.67]
    \chopper{tiled_PS/buddha_directlight_Ours_n1_box_perceptual_spectrum}
\end{tikzpicture}
\\

\begin{tikzpicture}[scale=0.67]
    \micro{tiled_PS/microcity_directlight_directlighting_n4_box_perceptual}
\end{tikzpicture}
&
\begin{tikzpicture}[scale=0.67]
    \micro{tiled_PS/microcity_directlight_Abdalla_n4_box_perceptual}
\end{tikzpicture}
&
\begin{tikzpicture}[scale=0.67]
    \micro{tiled_PS/microcity_directlight_Heitz21_n4_box_perceptual}
\end{tikzpicture}
&
\begin{tikzpicture}[scale=0.67]
    \micro{tiled_PS/microcity_directlight_Ours_n4_box_perceptual}
\end{tikzpicture}
\\
\begin{tikzpicture}[scale=0.67]
    \micro{tiled_PS/microcity_directlight_directlighting_n4_box_perceptual_spectrum}
\end{tikzpicture}
&
\begin{tikzpicture}[scale=0.67]
    \micro{tiled_PS/microcity_directlight_Abdalla_n4_box_perceptual_spectrum}
\end{tikzpicture}
&
\begin{tikzpicture}[scale=0.67]
    \micro{tiled_PS/microcity_directlight_Heitz21_n4_box_perceptual_spectrum}
\end{tikzpicture}
&
\begin{tikzpicture}[scale=0.67]
    \micro{tiled_PS/microcity_directlight_Ours_n4_box_perceptual_spectrum}
\end{tikzpicture}
\end{tabularx} 
     \vspace{-2.5mm}
     \caption{
        Comparison of rendering our method with \citet{abdalla2020screen} and \citet{belcour2021rank1}. The first line shows the rendering at 1 samples per pixel for the buddha scene and at 4 samples per pixel for the microcity scene. The second line shows the tilled power spectrum of the error with the reference.
     }
     \label{fig:tilled_PS_buddha}
 \end{figure*}
 \begin{figure*}[t!]
    \centering

\newcommand{\OneImage}[1]{
    \begin{scope}
        \clip (0,0)-- (0.0,4.65) -- (8.75,4.65) -- (8.75,0.0) -- cycle;
        \path[fill overzoom image=figures/#1] (0,0) rectangle (8.5,4.5);
    \end{scope}
}

\begin{tikzpicture}[scale=2.1]
        \OneImage{stipling/elephants.pdf}
\end{tikzpicture} 
     \vspace{-2.5mm}
     \caption{
         Stippling image of elephants in grey scale. The point-set is composed of 100 000 samples optimized for a single class problem.
     }
     \label{fig:stippling_elephants}
 \end{figure*}


\bibliographystyle{ACM-Reference-Format}
\bibliography{paper}